\documentclass[showpacs,twocolumn,nofootinbib]{revtex4}
\usepackage{amsmath}
\usepackage{graphicx}
\usepackage{amssymb}

\input{tcilatex}
\begin{document}

\title{Spin-dominated waveforms for unequal mass compact binaries}
\author{M\'{a}rton T\'{a}pai$^{\dag }$, Zolt\'{a}n Keresztes$^{\ddag }$, L%
\'{a}szl\'{o} \'{A}rp\'{a}d Gergely$^{\star }$}
\affiliation{Department of Theoretical Physics, University of Szeged, Tisza Lajos krt
84-86, Szeged 6720, Hungary\\
Department of Experimental Physics, University of Szeged, D\'{o}m t\'{e}r 9,
Szeged 6720, Hungary\\
$^{\dag }${\small E-mail: tapai@titan.physx.u-szeged.hu\quad }$^{\star }$%
{\small \ E-mail: gergely@physx.u-szeged.hu\quad }$^{\ddag }${\small E-mail:
zkeresztes@titan.physx.u-szeged.hu}}

\begin{abstract}
We derive spin-dominated waveforms (SDW) for binary systems composed of
spinning black holes with unequal masses (less than $1:30$). Such systems
could be formed by an astrophysical black hole with a smaller black hole or
a neutron star companion; and typically arise for supermassive black hole
encounters. SDW characterize the last stages of the inspiral, when the
larger spin dominates over the orbital angular momentum (while the spin of
the smaller companion can be neglected). They emerge as a double expansion
in the post-Newtonian parameter $\varepsilon $ and the ratio $\xi $ of the
orbital angular momentum and dominant spin. The SDW amplitudes are presented
to ($\varepsilon ^{3/2},\xi $) orders, while the phase of the gravitational
waves to ($\varepsilon ^{2},\xi $) orders (omitting the highest order mixed
terms). To this accuracy the amplitude includes the (leading order)
spin-orbit contributions, while the phase the (leading order) spin-orbit,
self-spin and mass quadrupole-monopole contributions. While the SDW hold for
any mass ratio smaller than $1:30$, lower bounds for the mass ratios are
derived from the best sensitivity frequency range expected for Advanced LIGO
(giving $1:140$), the Einstein Telescope ($7\times 10^{-4}$), the LAGRANGE ($%
7\times 10^{-7}$) and LISA missions ($7\times 10^{-9}$), respectively.
\end{abstract}

\pacs{04.25.Nx, 04.30.Db, 04.80.Nn, 95.85.Sz}
\maketitle

%\date{}

\section{Introduction}

The orbital evolution of the binary system forming compact objects is
accompanied by emission of gravitational waves. The slight dissipative
effect of the gravitational radiation allows for a long inspiral phase,
followed by a short merger/plunge \cite{plunge} and the relaxation of the
newly formed object in a process called ringdown \cite{ringdown}. The
inspiral dynamics can be accurately described by a post-Newtonian (PN)
expansion in terms of a small (but ever increasing) parameter $\varepsilon
=Gm/c^{2}r\approx v^{2}/c^{2}$ (with $m=m_{1}+m_{2}$ the total mass, $r$ the
orbital separation and$\ v$ the orbital velocity of the reduced mass
particle $\mu =m_{1}m_{2}/m$).

The inspiral dynamics including the contributions of the spins and
quadrupole moments was extensively discussed in Refs. \cite{BOC}-\cite%
{Inspiral2}. The gravitational waveforms $h_{+}$ and $h_{\times }$
characterizing the inspiral on circular orbits were given to 1 PN order in
Ref. \cite{KIDDER}, including the leading order spin-orbit contributions. In
Ref. \cite{ABFO} the spinning waveforms were given to 1.5 PN accuracy, and
specified for the equal mass case by an expansion in the small angle $\iota $
(which will be denoted $\alpha $ in our formalism, and will \textit{not} be
small) span by the orbital angular momentum $\mathbf{L}_{N}$ and total
angular momentum $\mathbf{J}$. Spinning gravitational waveforms on generic
orbits were also given in Ref. \cite{MajarVasuth1} to 1.5 PN order accuracy,
and corrected in Ref. \cite{CornishKey}. The multipole moments including
spin effects to 2.5 PN order for gravitational wave amplitudes were
calculated in Ref. \cite{multipoles}.

The mass range of neutron stars is relatively narrow due to the upper bound
on their masses represented by the Oppenheimer-Volkoff limit \cite{OV}, \cite%
{Bombacci}. Consequently the equal or nearly equal mass compact binary model
stands as a good approximation during their inspiral.

The situation is however radically different for black hole binaries, and
these are the compact objects, which have significant spin. There is no
reason to believe that for astrophysical black hole binaries (with each
black hole having a mass extending from a few solar masses, M$_{\odot }$ to
a few ten times of M$_{\odot }$) the comparable mass case is more likely
than any other mass ratio $\nu =m_{2}/m_{1}$. Moreover, for supermassive
black hole binaries (with masses $3\times 10^{6}-3\times 10^{9}$ M$%
_{\odot }$) it has been shown that a mass ratio between $0.3$ and $0.03$ is
typical \cite{SpinFlip1}, \cite{SpinFlip2}, \cite{MassSpin} equal mass and
extreme mass ratio encounters being disfavored. Low mass ratio and
intermediate mass black hole binaries emitting gravitational waves are also
believed to be sources for the LIGO detectors. However no signals were
detected yet \cite{LIGO1}, \cite{LIGO2}, \cite{LIGO3}.With the advent of
advanced LIGO \cite{aLIGO}, the planned space-born missions LISA-eLISA/NGO 
\cite{LISA}-\cite{eLISA} and LAGRANGE \cite{LAGRANGE}, also the third
generation detectors, like the Einstein Telescope \cite{ET}, the knowledge
of waveforms from binaries with smaller mass ratio ranges becomes
imperative. The small mass ratio parameter in these cases stands as a second
small parameter, which modifies (and simplifies) some of the immediate
results standing for the equal mass case. Indeed, it has been shown in Ref. \cite%
{SpinFlip1}, that the ratio of the spin magnitudes is%
\begin{equation}
\frac{S_{2}}{S_{1}}=\frac{\chi _{2}}{\chi _{1}}\nu ^{2}~,  \label{S2S1}
\end{equation}%
where $\chi _{i}\in \left[ 0,1\right] $ are the dimensionless spin
parameters. For rapidly spinning compact binaries and for small mass ratio
the role of the second spin becomes negligible.

Moreover, the ratio of the dominant spin and magnitude of Newtonian orbital
angular momentum is \cite{SpinFlip1}%
\begin{equation}
\frac{S_{1}}{L_{N}}\approx \varepsilon ^{1/2}\nu ^{-1}\chi _{1}~.
\label{S1LN}
\end{equation}%
This relation shows that while for equal masses the inspiral is dominated by 
$L_{N}$, for small mass ratios this can change. Indeed, $S_{1}$ becomes
dominant in the last stages of the inspiral for the unequal mass case $\nu
<0.1$. (Obviously, in the test particle limit $S_{1}$ dominates throughout
the inspiral, the orbital angular momentum of the test particle being
negligible.)

In this paper we will focus on the situation when $\nu $ is small and we
will derive the corresponding spinning waveforms to accuracy of order $\nu $%
. This immediately implies to disregard $S_{2}$ as of order $\nu ^{2}$
compared to $S_{1}$.\footnote{%
Gravitational waveforms with the inclusion of only one spin were previously
investigated in Refs. \cite{BCV2}, \cite{PBCV}, to leading order in
amplitude and 3.5 PN orders in the phase (the physical template family),
however omitting the quadrupole-monopole and self-spin contributions, given
later in Ref. \cite{omegadot}.} As the gravitational wave amplitude
increases drastically toward the end of the inspiral, we will concentrate
on this regime. As has been shown in \cite{SpinFlip1}, \cite{SpinFlip2}, in
this mass ratio range the end of the inspiral is characterized by $S_{1}$
dominating over $L_{N}.$ Hence we call the corresponding gravitational
waveforms as \textit{spin-dominated waveforms} (SDW). We express this
dominance by introducing a new small parameter 
\begin{equation}
\xi =\varepsilon ^{-1/2}\nu ~,  \label{xibev}
\end{equation}%
assuming $\xi \leq 0.1.$ As the PN parameter increases during the inspiral,
this condition selects the last part of the inspiral starting from a given
radius $r_{1}$. Hence the SDW are valid from $r_{1}$ to the end of
the validity of the PN expansion.\footnote{%
The mass ratio $\nu $ will be replaced by $\varepsilon ^{1/2}\xi $ in all
expressions, with the exception of leading order contributions, where we
keep $\nu $.}

We derive the SDW inspiral waveforms for circular orbits as a double
expansion in the parameters $\varepsilon $ and $\xi ,$ with a linear
accuracy in $\xi $ and $3/2$ orders in $\varepsilon $ \textbf{(}dropping
however the $\varepsilon ^{3/2}\xi $ terms\textbf{)}. The introduction of
the parameter $\xi $ leads to the natural neglection of the second spin,
since $S_{i}=G/cm_{i}^{2}\chi _{i}$ such that from Eq. (\ref{S2S1}) we find $%
S_{2}=\chi _{2}G/cm_{1}^{2}\nu ^{2}=\chi _{2}G/cm_{1}^{2}\varepsilon \xi
^{2} $. Since the first terms containing spins in the amplitude are the 1
PN\ order terms \cite{KIDDER}, the leading order terms containing $S_{2}$
would be shifted to 2PN orders in the amplitude, to be neglected in our
approach (unless $\xi \gg 1$, which is not what we assume in this paper).

The structure of the paper is as follows. In Sec. \ref{Approx} we analyze
the limits of validity of the proposed SDW. In Sec. \ref{SOWF}
and Appendix \ref{appA}\ we derive and present the SDW (gravitational wave
amplitudes, with the spin-orbit corrections included) resulting from the
double expansion. Here we employ that in the discussed parameter region the
angle $\beta _{1}$ span by the dominant spin and total angular momentum is
small, of order $\xi $. Therefore wherever possible, we express the angle $%
\alpha $ span by the total angular momentum and orbital angular momentum in
terms of $\beta _{1}$ and perform the respective expansions to first order
in $\xi $. Section \ref{fazis} contains the expressions of the gravitational
wave phase, expanded in terms of both $\varepsilon $ and $\xi $, with a
linear accuracy in $\xi $ and $\,$second order in $\varepsilon $ \textbf{(}%
dropping however the $\varepsilon ^{2}\xi $ terms\textbf{)}. The phase then
contains spin effects (the leading order spin-orbit and self-spin) and mass
quadrupole-monopole effects, however spin-spin terms (of $\varepsilon
^{5/2}\xi $ order) were dropped, as they would be of the same order in $%
\varepsilon $ as the PN\ correction of the spin-orbit terms \cite{BFB}, not
included in our treatment.

Finally we present our concluding remarks in Sec. \ref{Conclusion}.

\section{Limits of validity\label{Approx}}

During the inspiral, the PN parameter increases. Therefore the condition $%
\xi \leq \xi _{1}=0.1$ cannot be obeyed at distances larger than a certain $%
r_{1}$, which depends on the particular mass ratio. The value of the
corresponding limiting PN parameter is $\varepsilon
_{1}=Gm/c^{2}r_{1}=100\nu ^{2}$. Starting from these values of the
separation and PN\ parameter, the approximation holds until the PN expansion
breaks down. Levin, McWilliams, and Contreras argue in Ref. \cite{LEVIN},
that this should be at $\varepsilon _{2}=0.1$, and we adopt this upper
limit. The argument goes as follows. At about $\varepsilon _{2}=0.1$ (well
above the innermost marginally stable circular orbit) the PN expansion
breaks down, as the 3.5 PN dissipation term (which is positive) becomes
comparable to the 2.5 PN order contribution (which is negative). Summing up,
the regime we study lies between $\varepsilon _{1}$ and $\varepsilon _{2}$.
With increasing mass ratio this interval shrinks, and it vanishes at $\nu
_{\max }=0.0316\approx 1:32$.

\begin{figure*}[tbp]
\includegraphics[height=10cm]{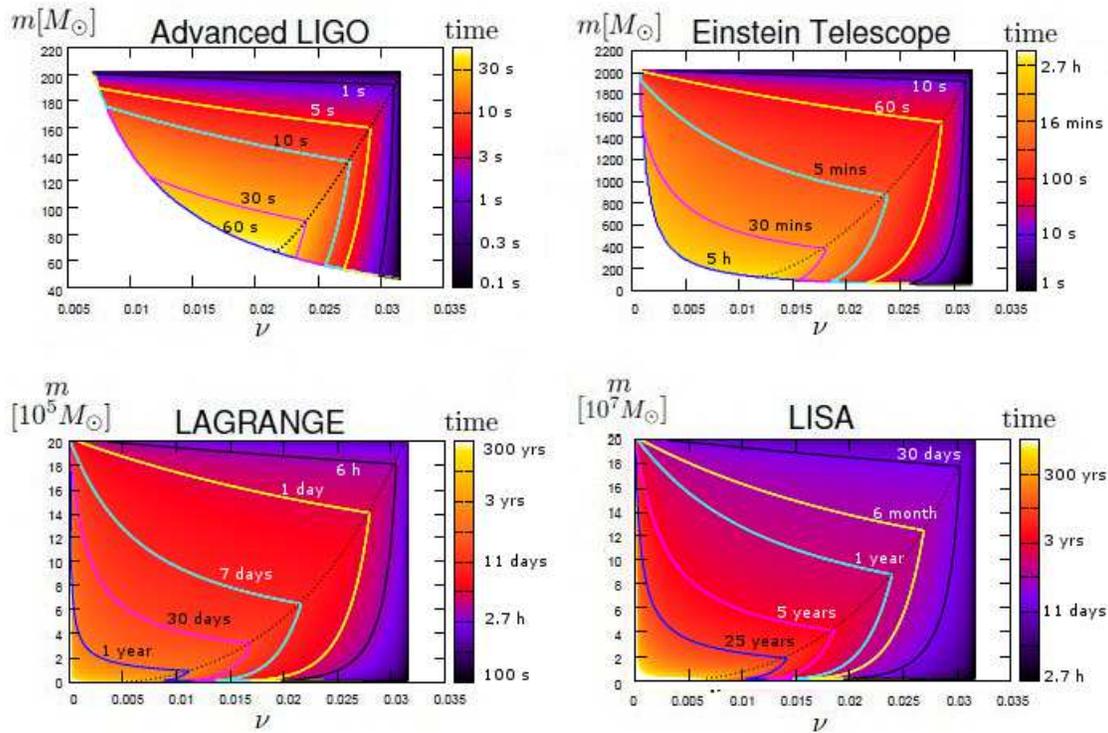}
\caption{(Color online) The color code and contour lines represent the time interval during
which a system evolves from $\max \left( \protect\varepsilon _{1},\protect%
\varepsilon _{f\min }\right) $ to $\protect\varepsilon _{2}$ for Advanced
LIGO (top left), Einstein Telescope (top right), LAGRANGE (bottom left), and
LISA (bottom right) gravitational wave detectors, represented as function of
the total mass $m$ and mass ratio $\protect\nu $. The color code and contour lines are displayed logarithmically. The upper bound for the total mass emerges
from the lower frequency limit of the best sensitivity range for each
detector. The (total mass dependent) minimal mass ratio (visible on the top
panels) arises by assuming the smaller mass to represent a neutron star with
typical mass of $1.4$ M$_{\odot }$. The maximal mass ratio (the cutoff on
the right) is determined by the requirement, that $\protect\varepsilon %
_{1}=100\protect\nu ^{2}\leq 0.1.$ In the larger part of the allowed
parameter space (the region on the left of each dashed curve, representing $%
\protect\varepsilon _{1}=\protect\varepsilon _{f\min }$) the gravitational
waves entering into the best sensitivity range of the respective detectors
are described by SDW. In the regions on the right of the dashed
curves however SDW can not be applied from the lower frequency bound of the
best sensitivity range, hence the time intervals are shorter.}
\label{Fig02}
\end{figure*}

For a specific gravitational wave detecting instrument, the lower frequency
limit of the best sensitivity band and $\varepsilon _{2}$ determines the
highest possible total mass of a binary source for the waves to be detected
by the respective instrument. Kepler's third law gives a leading order
estimate to the total mass $m$ of the system on circular orbit in terms of
the gravitational wave frequency $f$ and PN parameter $\varepsilon $ as 
\textbf{\ }%
\begin{equation}
m=\frac{c^{3}}{\pi G}\varepsilon ^{3/2}f^{-1}~.  \label{epsomega}
\end{equation}%
The lower frequency bounds are $10$ Hz for Advanced LIGO \cite{aLIGO}, $1$
Hz for the Einstein Telescope \cite{ET}, $10^{-3}$ Hz for LAGRANGE \cite%
{LAGRANGE} and $10^{-5}$ Hz for LISA \cite{LISA}. The respective values for $%
m_{\max }$ are thus $202M_{\odot }$ (Advanced LIGO), $2020M_{\odot }$
(Einstein Telescope), $2\times 10^{6}M_{\odot }$ (LAGRANGE) and $2\times
10^{8}M_{\odot }$ (LISA).

Further, a lower limit $\nu _{\min }$ arises if we fix the smaller mass to
be that of a neutron star. This leads to $\nu _{\min }\approx
0.00\allowbreak 7\approx 1:143$ for Advanced LIGO, $\nu _{\min }\approx
7\times 10^{-4}$ for Einstein Telescope, $\nu _{\min }\approx 7\times
10^{-7}~$for LAGRANGE and $\nu _{\min }\approx 7\times 10^{-9}$ for LISA.

Next we estimate the time-interval spent by the binary evolving (due to
gravitational radiation dissipation, considered to leading order) on
quasicircular orbits from $\varepsilon _{1}$ to $\varepsilon _{2}$, as
function of the mass ratio $\nu \in \left[ \nu _{\min ,}\nu _{\max }\right] $%
. For this we rewrite the leading order radiative orbital angular frequency
evolution given in Ref. \cite{omegadot} as%
\begin{equation}
\frac{(1+\nu )^{2}}{\nu }\omega ^{-8/3}=\frac{2^{8}\left( Gm\right) ^{5/3}}{%
5c^{5}}(t_{c}-t)~,
\end{equation}%
(where $t_{c}-t$ is the time left until the final coalescence). Rewriting
Eq. (\ref{epsomega}) as%
\begin{equation}
\varepsilon =\frac{\left( Gm\omega \right) ^{2/3}}{c^{2}}~,
\end{equation}%
where $\omega =\pi f$, we obtain%
\begin{equation}
\frac{(1+\nu )^{2}}{\nu }\varepsilon ^{-4}=\frac{2^{8}c^{3}}{5Gm}(t_{c}-t)~.
\end{equation}%
Then the time elapsed during the evolution from $\varepsilon _{1}$ to $%
\varepsilon _{2}$ can be calculated as 
\begin{equation}
\Delta t=\frac{5Gm}{2^{8}c^{3}}\frac{(1+\nu )^{2}}{\nu }\left( \varepsilon
_{1}^{-4}-\varepsilon _{2}^{-4}\right) ~.  \label{deltatimeform}
\end{equation}

Thinking of a specific instrument we can investigate the time necessary for
the evolution starting from the lower frequency range ($f_{\min }$) of a
given detector (when this belongs to a value $\varepsilon _{f\min
}>\varepsilon _{1}$) until $\varepsilon _{2}$. On Fig. \ref{Fig02} we
represent this time interval necessary for the evolution from $\max \left(
\varepsilon _{1},\varepsilon _{f\min }\right) $ to $\varepsilon _{2}$ for
the enlisted detectors (Advanced LIGO, Einstein Telescope, LAGRANGE, and
LISA).

\section{Spin-dominated gravitational waveforms\label{SOWF}}

\subsection{Gravitational waveforms to 1.5PN order}

Gravitational waveforms including spin-orbit effects were previously
calculated in Ref.\ \cite{KIDDER} to 1.5 PN order for nearly circular
orbits, based on the radiative multipole moments of 1.5 PN order accuracy,
and reproduced (with some misprints corrected) in Ref. \cite{ABFO} as%
\begin{eqnarray}
h^{ij} &=&{\frac{2G\mu }{c^{4}D}}\left( {\frac{Gm}{r}}\right) \biggl[%
Q_{c}^{ij}  \notag \\
&&+P^{0.5}Q_{c}^{ij}\varepsilon ^{1/2}+PQ_{c}^{ij}\varepsilon   \notag \\
&&+\left( P^{1.5}Q_{c}^{ij}+P^{1.5}Q_{Tail}^{ij}\right) \varepsilon ^{3/2}%
\biggr]_{TT~}~,  \label{hij}
\end{eqnarray}%
where TT denotes the transverse trace-free projection into the plane
orthogonal to the direction $\mathbf{\hat{N}}$ pointing from the source to
the observer (the line of sight), and

\begin{equation}
Q_{c}^{ij}=2\left[ \lambda ^{i}\lambda ^{j}-\hat{r}^{i}\hat{r}^{j}\right] ~,
\end{equation}%
\begin{eqnarray}
P^{0.5}Q_{c}^{ij} &=&{\frac{1-\nu }{1+\nu }}\left\{ 6(\mathbf{\hat{N}\cdot 
\hat{r}})\hat{r}^{(i}\lambda ^{j)}\right.  \notag \\
&&\left. +(\mathbf{\hat{N}\cdot \hat{\lambda}})\left[ \hat{r}^{i}\hat{r}%
^{j}-2\lambda ^{i}\lambda ^{j}\right] \right\} ~,
\end{eqnarray}%
\begin{eqnarray}
PQ_{c}^{ij} &=&{\frac{2}{3}}(1-3\eta )\biggl\{(\mathbf{\hat{N}\cdot \hat{r}}%
)^{2}\left[ 5\hat{r}^{i}\hat{r}^{j}-7\lambda ^{i}\lambda ^{j}\right]  \notag
\\
&&-16(\mathbf{\hat{N}\cdot \hat{n}})(\mathbf{\hat{N}\cdot \hat{\lambda}})%
\hat{r}^{(i}\lambda ^{j)}  \notag \\
&&\mbox{}+(\mathbf{\hat{N}\cdot \hat{\lambda}})^{2}\left[ 3\lambda
^{i}\lambda ^{j}-\hat{r}^{i}\hat{r}^{j}\right] \biggr\}  \notag \\
&&+{\frac{1}{3}}(19-3\eta )(\hat{r}^{i}\hat{r}^{j}-\lambda ^{i}\lambda ^{j})
\notag \\
&&\mbox{}+{\frac{2c}{Gm^{2}}}\frac{1+\nu }{1-\nu }\hat{r}^{(i}(\mathbf{%
(\sigma -S)\times \hat{N}})^{j)}~,
\end{eqnarray}%
\begin{eqnarray}
P^{1.5}Q_{c}^{ij} &=&{\frac{1-\nu }{1+\nu }}\biggl\{(1-2\eta )\biggl[{\frac{1%
}{2}}(\mathbf{\hat{N}\cdot \hat{\lambda}})^{3}\left( \hat{r}^{i}\hat{r}%
^{j}-4\lambda ^{i}\lambda ^{j}\right)  \notag \\
&&+{\frac{1}{4}}(\mathbf{\hat{N}\cdot \hat{r}})^{2}(\mathbf{\hat{N}\cdot 
\hat{\lambda}})\left( 58\lambda ^{i}\lambda ^{j}-37\hat{r}^{i}\hat{r}%
^{j}\right)  \notag \\
&&\mbox{}-{\frac{65}{6}}(\mathbf{\hat{N}\cdot \hat{r}})^{3}\hat{r}%
^{(i}\lambda ^{j)}+15(\mathbf{\hat{N}\cdot \hat{r}})(\mathbf{\hat{N}\cdot 
\hat{\lambda}})^{2}\hat{r}^{(i}\lambda ^{j)}\biggr]  \notag \\
&&-(\mathbf{\hat{N}\cdot \hat{\lambda}})\biggl[{\frac{1}{12}}(101-12\eta )%
\hat{r}^{i}\hat{r}^{j}\mbox{}-{\frac{1}{2}}(19-4\eta )\lambda ^{i}\lambda
^{j}\biggr]  \notag \\
&&-{\frac{1}{6}}(149-36\eta )(\mathbf{\hat{N}\cdot \hat{r}})\hat{r}%
^{(i}\lambda ^{j)}\biggr\}  \notag \\
&&\mbox{}-{\frac{2c}{m^{2}G}}\Biggl\{\lambda ^{i}\lambda ^{j}\left[ \mathbf{%
\hat{L}_{N}\cdot }(2\mathbf{S}+3\mathbf{\sigma })\right]  \notag \\
&&\mbox{}-6\hat{r}^{i}\hat{r}^{j}\left[ \mathbf{\hat{L}_{N}\cdot }(\mathbf{S}%
+\mathbf{\sigma })\right]  \notag \\
&&+2\lambda ^{(i}\left[ \mathbf{\hat{r}\times }(\mathbf{\sigma })\right]
^{j)}+\hat{r}^{(i}\left[ \mathbf{\hat{\lambda}\times }(4\mathbf{S}+5\mathbf{%
\sigma })\right] ^{j)}  \notag \\
&&\mbox{}+2(\mathbf{\hat{N}\cdot \hat{\lambda}})\left[ (\mathbf{\sigma })%
\mathbf{\times \hat{N}}\right] ^{(i}\hat{r}^{j)}  \notag \\
&&+2(\mathbf{\hat{N}\cdot \hat{r}})\left[ (\mathbf{\sigma })\mathbf{\times 
\hat{N}}\right] ^{(i}\lambda ^{j)}\Biggr\}~,
\end{eqnarray}%
\begin{eqnarray}
P^{1.5}Q_{Tail}^{ij} &=&4\left[ \pi (\lambda ^{i}\lambda ^{j}-\hat{r}^{i}%
\hat{r}^{j})\right.  \notag \\
&&\left. +4\ln \left( \frac{\omega }{\omega _{0}}\right) \hat{r}^{(i}\lambda
^{j)}\right] ~.
\end{eqnarray}%
Here $r^{i}$ and $\lambda ^{i}$ are the components of the separation vector
and of $\mathbf{\lambda }=\mathbf{L}_{\mathbf{N}}\times \mathbf{r=v/}r\omega 
$, the parameter $\eta =\mu /m=\nu /\left( 1+\nu \right) ^{2}$\ is the
symmetric mass ratio\textbf{, }and we introduced the spin combinations $%
\mathbf{S}=\mathbf{S}_{1}+\mathbf{S}_{2}$ and $\mathbf{\sigma }=\nu \mathbf{S%
}_{1}+\nu ^{-1}\mathbf{S}_{2}$. The last term, $P^{1.5}Q_{Tail}^{ij}$
arising from the contributions of the gravitational wave tails was first
presented in this form in Ref. \cite{ABFO}. The parameter $\omega _{0}$ is
an arbitrary constant frequency scale \cite{BIWW}. Unit vectors carry a hat.

The two gravitational wave polarization states $h_{+}$ and $h_{\times }$
emerge as the linear combinations 
\begin{eqnarray}
h_{+} &=&\frac{1}{2}\left( \mathbf{\hat{x}}_{i}\mathbf{\hat{x}}_{j}-\mathbf{%
\hat{y}}_{i}\mathbf{\hat{y}}_{j}\right) h^{ij}~,  \label{hplusgen} \\
h_{\times } &=&\frac{1}{2}\left( \mathbf{\hat{x}}_{i}\mathbf{\hat{y}}_{j}+%
\mathbf{\hat{x}}_{j}\mathbf{\hat{y}}_{i}\right) h^{ij}~.  \label{hcrossgen}
\end{eqnarray}%
of the $h^{ij}$ tensor components perpendicular to $\mathbf{\hat{N}}$ \cite%
{KIDDER}, with $\mathbf{\hat{y}}=\mathbf{J}\times \mathbf{\hat{N}/}%
\left\vert \mathbf{J}\times \mathbf{\hat{N}}\right\vert $ and $\mathbf{\hat{x%
}}=\mathbf{\hat{y}\times \hat{N}}$\footnote{%
We mention that the polarization vectors $\mathbf{\hat{x}}$ and $\mathbf{%
\hat{y}}$ are rotated by $\pi /2$ in Ref. \cite{ABFO}, resulting a global
sign difference in the definition of $h_{+}$ and $h_{\times }$.}.

\subsection{The spin-dominated regime}

The total angular momentum $\mathbf{J}=\mathbf{L}_{\mathbf{N}}+\mathbf{L}_{%
\mathbf{PN}}+\mathbf{L}_{\mathbf{SO}}+\mathbf{S}_{\mathbf{1}}+\mathbf{S}_{%
\mathbf{2}}$ is conserved to 2PN accuracy \cite{KWW}. Here $\mathbf{L}_{%
\mathbf{PN}}=\epsilon _{PN}\mathbf{L}_{\mathbf{N}}$ and $\mathbf{L}_{\mathbf{%
SO}}$ are the PN and spin-orbit corrections in the orbital angular momentum
vector. Starting from their expressions Eq. (39) in Ref. \cite{Inspiral1}
and Eq. (B23) in Ref.\cite{Inspiral2}, and the definition of the PN\
parameter $\varepsilon ~$we find

\begin{eqnarray}
\frac{L_{PN}}{L_{N}} &=&\epsilon _{PN}\approx \varepsilon \left[ \frac{7}{2}-%
\frac{1}{2}\frac{\xi \varepsilon ^{1/2}}{(1+\xi \varepsilon ^{1/2})^{2}}%
\right]  \label{epspn} \\
\frac{L_{SO}}{L_{N}} &\approx &\frac{\xi ^{2}\varepsilon ^{5/2}}{4\left(
1+\xi \varepsilon ^{1/2}\right) ^{4}}\left[ 4\xi ^{-1}\varepsilon
^{-1/2}\chi _{1}\right.  \notag \\
&&\left. +3\left( \chi _{1}+\chi _{2}\right) +4\xi \varepsilon ^{1/2}\chi
_{2}\right] ~.
\end{eqnarray}%
Expanding these to linear order in $\xi $, we have%
\begin{eqnarray}
\frac{L_{PN}}{L_{N}} &=&\mathcal{O}\left( \varepsilon ,\varepsilon ^{3/2}\xi
\right) ~,  \notag \\
\frac{L_{SO}}{L_{N}} &=&\mathcal{O}\left( \varepsilon ^{2}\xi \chi
_{1}\right) ~.
\end{eqnarray}%
Keeping the terms to $\varepsilon ,~\varepsilon \xi ,~\varepsilon ^{3/2}$
orders only in the above set of equations, $\mathbf{J}$ becomes%
\begin{equation}
\mathbf{J}=\left( 1+\frac{7}{2}\varepsilon \right) \mathbf{L}_{\mathbf{N}}+%
\mathbf{S}_{\mathbf{1}}~.  \label{Jdecomp}
\end{equation}%
Therefore to the accuracy we are interested in, both the second spin and the
spin-orbit contribution to the orbital angular momentum could be dropped.
Then in the triangle with sides $S_{1}$, $\left( 1+7\varepsilon /2\right)
L_{N}$ and $J$, containing the angles $\alpha =\arccos \left( \mathbf{\hat{J}%
\cdot \hat{L}}_{\mathbf{N}}\right) $ and $\beta _{1}=\arccos \left( \mathbf{%
\hat{J}\cdot \hat{S}}_{\mathbf{1}}\right) $ the sine theorem gives 
\begin{eqnarray}
\sin \beta _{1} &=&\left( 1+\frac{7}{2}\varepsilon \right) \frac{L_{N}}{S_{1}%
}\sin \alpha ~  \notag \\
&=&\left( 1+\frac{7}{2}\varepsilon \right) \frac{\xi }{\chi _{1}}\sin \alpha
~.
\end{eqnarray}%
As $\alpha $ has no preferred value and $\xi $ is small, we conclude that $%
\beta _{1}$ is of order $\xi $. (From the assumption $\xi <0.1$ it follows
that $\sin \beta _{1}\lesssim 0.1$.) Thus any trigonometric function of $%
\beta _{1}$ can be approximated by its Taylor expanded form, to first order
accuracy (dropping $\xi ^{2}\leq 0.01$ terms and higher).

\begin{figure}[th]
%\begin{center}
\includegraphics[height=6cm]{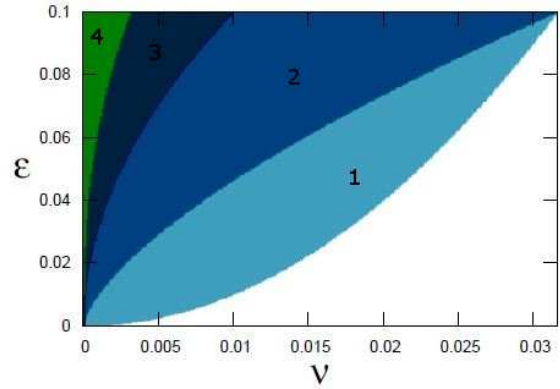} %\end{center}
\caption{(Color online) The figure represents the parameter ranges $\protect%
\nu $ and $\protect\varepsilon $ for which all terms given in the waveform
are larger than the pure 2PN ($\protect\varepsilon ^{2}$ ) contributions
(area 1). In the parameter range 2 the contributions $\protect\varepsilon 
\protect\xi ,~\protect\varepsilon \protect\beta _{1}$ are becoming
comparable with the pure 2PN terms, falling outside of the limits of
validity of our approach, hence they can be neglected. In the parameter
range 3 all mixed terms in the double expansion become negligibly small,
while in the parameter range 4 all terms containing $\protect\xi $ and $%
\protect\beta _{1}$ can be dropped. In the white area, the condition $%
\protect\xi \leq 0.1$ doesn't hold during the inspiral.}
\label{Fig3}
\end{figure}

It was shown in Ref. \cite{SpinFlip1}, that the angle $\kappa _{1}=\alpha
+\beta _{1}$ stays constant during the inspiral, when the smaller spin is
negligible. Therefore we replace $\alpha $ everywhere by $\kappa _{1}-\beta
_{1}$ and then expand to first order in $\beta _{1}$. We will show in the
next subsection that the double expansion of $h_{+}$ and $h_{\times }$ in $%
\varepsilon $ and $\xi $ leads to the following structure of terms: $%
1,~\beta _{1},~\varepsilon ^{1/2},~\varepsilon ^{1/2}\beta _{1},~\varepsilon
,~\varepsilon \xi ,~\varepsilon ^{3/2}$. (We note here that the leading
order contribution from the smaller spin $S_{2}$ would appear only at $%
\varepsilon ^{2}\xi ^{2}$ orders.) Since $\varepsilon $ increases during the
inspiral, the terms $\beta _{1},~\varepsilon ^{1/2}\beta _{1},~\varepsilon
\xi $, all kept, increase as well, however the $\varepsilon ^{2}$ term,
discarded, increases at an even faster rate. Therefore when either of the
enlisted terms becomes comparable to the $\varepsilon ^{2}$ contribution, we
have to drop it as well. Figure \ref{Fig3} shows the terms to be kept
in the parameter space $\left( \varepsilon ,\nu \right) $. 

\subsection{SDW inspiral waveforms}

We give the gravitational wave polarizations (\ref{hplusgen}) and (\ref%
{hcrossgen}) in the source coordinate system defined as follows. The $z_{S}$%
-axis lies along the direction of $\mathbf{J}$\textbf{,} and the $x_{S}$%
-axis is assigned by the projection of $\mathbf{\hat{N}}$ perpendicular to $%
\mathbf{J}$. The polar angle of $\mathbf{\hat{N}}$ will be denoted by $%
\theta $. In this coordinate system the components of the vectors $\mathbf{r}
$, $\mathbf{\lambda }$, and $\mathbf{S}_{1}$ are

\begin{eqnarray}
\frac{\mathbf{r}_{x}}{r} &=&-\sin \left( \frac{3\pi }{2}-\phi _{n}\right)
\cos \phi  \notag \\
&&-\cos \alpha \cos \left( \frac{3\pi }{2}-\phi _{n}\right) \sin \phi ~, 
\notag \\
\frac{\mathbf{r}_{y}}{r} &=&\cos \left( \frac{3\pi }{2}-\phi _{n}\right)
\cos \phi  \notag \\
&&-\cos \alpha \sin \left( \frac{3\pi }{2}-\phi _{n}\right) \sin \phi ~, 
\notag \\
\frac{\mathbf{r}_{z}}{r} &=&\sin \alpha \sin \phi ~,  \label{rcomp}
\end{eqnarray}%
\begin{eqnarray}
\mathbf{\lambda }_{x} &=&\sin \left( \frac{3\pi }{2}-\phi _{n}\right) \sin
\phi  \notag \\
&&-\cos \alpha \cos \left( \frac{3\pi }{2}-\phi _{n}\right) \cos \phi ~, 
\notag \\
\mathbf{\lambda }_{y} &=&-\cos \left( \frac{3\pi }{2}-\phi _{n}\right) \sin
\phi  \notag \\
&&-\cos \alpha \sin \left( \frac{3\pi }{2}-\phi _{n}\right) \cos \phi ~, 
\notag \\
\mathbf{\lambda }_{y} &=&\sin \alpha \cos \phi ~,  \label{lambdacomp}
\end{eqnarray}%
\begin{eqnarray}
S_{1x} &=&\frac{G}{c}m^{2}\eta \nu ^{-1}\chi _{1}\sin \beta _{1}\cos \left(
\phi _{1}-\phi _{n}\right) ~,  \notag \\
S_{1y} &=&\frac{G}{c}m^{2}\eta \nu ^{-1}\chi _{1}\sin \beta _{1}\sin \left(
\phi _{1}-\phi _{n}\right) ~,  \notag \\
S_{1z} &=&\frac{G}{c}m^{2}\eta \nu ^{-1}\chi _{1}\cos \beta _{1}~.
\label{S1comp}
\end{eqnarray}

Here the angle $\phi _{1}$ is the azimuthal angle of the dominant spin. $%
\phi _{n}$ is defined as the angle between the intersection of the planes
perpendicular to $\mathbf{J}$\ and $\mathbf{L}_{\mathbf{N}}$ and an
arbitrary inertial axis taken in the plane perpendicular to $\mathbf{J}$,
and $\phi $ is the orbital phase. These angles are shown on Figs. 1 and 2 of
Ref. \cite{Inspiral1}. The time evolution of these angles was given in Ref. 
\cite{Inspiral2}.

We substitute $\mathbf{r}$, $\mathbf{\lambda }$, and $\mathbf{S}_{1}$ from
Eqs. (\ref{rcomp})-(\ref{S1comp}) into the expression of $h_{+}$ and $%
h_{\times }$ given by Eqs. (\ref{hplusgen}) and (\ref{hcrossgen}). The 1.5
PN tail contribution contains terms with $\ln \left( \omega /\omega
_{0}\right) $, which can be absorbed into the leading order term by
redefining the phase $\phi $ as $\psi =\phi -2\varepsilon ^{3/2}\ln \left(
\omega /\omega _{0}\right) $\ called shifted phase \cite{BIWW}, \cite{ABFO}.
The change of $\phi $ into $\psi $ in the higher order terms gives
modification to the waveforms, however these are of higher order than $%
\varepsilon ^{3/2}$.\textbf{\ }The double expansion, including the redshift
dependence, of $h_{+}$ and $h_{\times }$ in $\varepsilon $ and $\xi $ leads
to the following structure to $\xi ~$and $\varepsilon ^{3/2}$ orders:

\begin{eqnarray}
h_{_{\times }^{+}} &=&\frac{2G^{2}m^{2}\varepsilon ^{1/2}\xi }{c^{4}Dr}\left[
h_{_{\times }^{+}}^{0}+\beta _{1}h_{_{\times }^{+}}^{0\beta }\right.   \notag
\\
&&+\varepsilon ^{1/2}\left( h_{_{\times }^{+}}^{0.5}+\beta _{1}h_{_{\times
}^{+}}^{0.5\beta }-2\xi h_{_{\times }^{+}}^{0}\right)   \notag \\
&&+\varepsilon \left( h_{_{\times }^{+}}^{1}-4\xi h_{_{\times
}^{+}}^{0.5}\right.   \notag \\
&&\left. +\beta _{1}h_{_{\times }^{+}}^{1\beta }+h_{_{\times
}^{+}}^{1SO}+\beta _{1}h_{_{\times }^{+}}^{1\beta SO}\right)   \notag \\
&&\left. +\varepsilon ^{3/2}\left( h_{_{\times }^{+}}^{1.5}+h_{_{\times
}^{+}}^{1.5SO}+h_{_{\times }^{+}}^{1.5tail}\right) \right] ~.
\label{waveform}
\end{eqnarray}

The indices of the various contributions to the waveform refer to the
respective order in the double expansion $\left( \varepsilon ,\xi \right) $,
respectively $\beta _{1}=\mathcal{O}\left( \xi \right) $. Thus the terms $%
h_{_{\times }^{+}}^{0}$ include leading order $\left( \varepsilon ^{0},\xi
^{0}\right) $ contributions, while the $h_{_{\times }^{+}}^{0\beta }$\ terms
are the $\left( \varepsilon ^{0},\beta _{1}\right) $ order contributions.
The rest of the terms without explicit spin magnitude dependence are $%
h_{_{\times }^{+}}^{0.5}$, of $\left( \varepsilon ^{0.5},\xi ^{0}\right) $
orders, $h_{_{\times }^{+}}^{0.5\beta }$, of $\left( \varepsilon
^{0.5},\beta _{1}\right) $ orders,  $h_{_{\times }^{+}}^{1}$, of $\left(
\varepsilon ^{1},\xi ^{0}\right) $\ orders, $h_{_{\times }^{+}}^{1\beta }$,
of $\left( \varepsilon ^{1},\beta _{1}\right) $\ orders and $h_{_{\times
}^{+}}^{1.5}$, of $\left( \varepsilon ^{1.5},\xi ^{0}\right) $\ orders respectively. The leading order contribution including the spin magnitude is 
$h_{_{\times }^{+}}^{1SO}$ of $\left( \varepsilon ^{1},\xi ^{0}\right) $
orders, and its corrections $h_{_{\times }^{+}}^{1\beta SO}$ and $%
h_{_{\times }^{+}}^{1.5SO}$ of $\left( \varepsilon ^{1},\beta _{1}\right) $
and $\left( \varepsilon ^{1.5},\xi ^{0}\right) $\ orders, respectively. The
leading order tail term is given by $h_{_{\times }^{+}}^{1.5tail}$, which is of $%
\left( \varepsilon ^{1.5},\xi ^{0}\right) $\ orders. All these explicit
expressions are given in the Appendix.

\section{The phase of the gravitational waveform in the spin dominated regime
\label{fazis}}

The radiative evolution of the orbital angular velocity up to 2PN orders,
including spin-orbit and spin-spin effects was given in \cite{KIDDER}.
Later, this was complemented by self-spin, quadrupole-monopole\footnote{%
The coefficient p\_i of Ref. \cite{omegadot} is $p_{i}=-G^{2}/c^{4}\omega
\chi _{i}^{2}m_{i}^{2}/m^{2}$, where $\omega =1$\ for black holes.} and
magnetic dipole-magnetic dipole contributions in \cite{omegadot}. Higher
order spin contributions were discussed in Refs. \cite{BFB}, \cite{Renorm}.

In order to employ these results for the SDWaveform, we need to derive the $%
\omega \left( \varepsilon \right) $ dependence to $\varepsilon ^{2}$ orders
accuracy.

Eq. (3) of Ref. \cite{omegadot} (leaving out the magnetic dipole-magnetic
dipole contributions, which are only important for magnetar binaries) gives: 
\begin{eqnarray}
\omega &=&\frac{\varepsilon ^{3/2}c^{3}}{Gm}\left\{ 1+\frac{3}{2}\varepsilon
+\left( -\frac{\xi }{2}+\chi _{1}\cos \kappa _{1}\right) \varepsilon
^{3/2}\right.  \notag \\
&&\left. \times \left[ \frac{171}{8}-\chi _{1}^{2}\left( -\frac{3}{8}+\frac{9}{8}%
\cos ^{2}\kappa _{1}\right) \right] \varepsilon ^{2}\right\}
\label{omegaeps}
\end{eqnarray}%
Note that the spin-spin terms are shifted to $\varepsilon ^{5/2}\xi $\
order, thus are neglected.

The radiative orbital angular velocity evolution ($\dot{\omega}$) is given
by Eq. (7) of Ref. \cite{omegadot}. This evolution for unequal mass case at
linear accuracy in $\xi $ and $\varepsilon ^{2}$ (with the neglection of $%
\varepsilon ^{2}\xi $) becomes%
\begin{eqnarray}
\dot{\omega} &=&\frac{96}{5}\frac{\varepsilon ^{6}\xi c^{6}}{\left(
Gm\right) ^{2}}\left\{ 1+\frac{1105}{336}\varepsilon \right.  \notag \\
&&+\left( 4\pi -\frac{79}{12}\xi -\frac{23}{4}\chi _{1}\cos \kappa
_{1}\right) \varepsilon ^{3/2}  \notag \\
&&+\left[ \frac{697\,465}{9072}\right.  \notag \\
&&\left. \left. +\chi _{1}^{2}\left( \allowbreak \frac{325}{96}\sin
^{2}\kappa _{1}-\allowbreak \frac{35}{16}\right) \right] \varepsilon
^{2}\right\} ~.
\end{eqnarray}%
Integrating Eq. (7) of Ref. \cite{omegadot} and employing Eq. (\ref{omegaeps}%
) we find to linear accuracy in $\xi $ and to $\varepsilon ^{2}$ orders
beyond leading order:%
\begin{eqnarray}
\tau &=&\frac{\varepsilon ^{-4}}{256}\left\{ 1-\frac{\allowbreak \allowbreak
265}{252}\varepsilon \right.  \notag \\
&&+\left[ 5\xi -\frac{32}{5}\pi +\frac{62}{5}\chi _{1}\cos \kappa _{1}\right]
\varepsilon ^{3/2}  \notag \\
&&+\left[ -\frac{24\,804\,463}{508\,032}\right.  \notag \\
&&\left. \left. +\chi _{1}^{2}\left( \frac{63}{8}-\frac{577}{48}\sin
^{2}\kappa _{1}\right) \right] \varepsilon ^{2}\right\} ~.  \label{taueps}
\end{eqnarray}%
Here $\tau $ is a dimensionless time variable defined as%
\begin{equation}
\tau =\frac{c^{3}}{5Gm}\left( 1-2\varepsilon ^{1/2}\xi \right) \left(
t_{c}-t\right) \ ,
\end{equation}%
where $t_{c}$ is the phase and the time at the final coalescence.
Integrating twice Eq. (7) of Ref. \cite{omegadot} gives the accumulated
orbital phase as 
\begin{eqnarray}
\phi _{c}-\phi &=&\frac{\varepsilon ^{-3}}{32\xi }\left\{ 1+2\varepsilon
^{1/2}\xi +\frac{1195}{1008}\varepsilon \right.  \notag \\
&&+\left( \allowbreak -10\pi +\frac{3925}{504}\xi +\frac{175}{8}\chi
_{1}\cos \kappa _{1}\right) \varepsilon ^{3/2}  \notag \\
&&+\left[ -\frac{21\,440\,675}{1016\,064}\right.  \notag \\
&&\left. \left. +\chi _{1}^{2}\left( \frac{375}{16}-\allowbreak \frac{3425}{%
96}\sin ^{2}\kappa _{1}\right) \right] \varepsilon ^{2}\allowbreak \right\}
~,
\end{eqnarray}%
where $\phi _{c}$ is the phase at the final coalescence$\allowbreak
\allowbreak .$

\section{Concluding Remarks\label{Conclusion}}

In this paper we have analyzed spinning compact binary inspiral in the case
when one of the binary components has much larger mass than the other one
and it spins fast. In such cases the end of the inspiral is characterized by
a much larger spin of the dominant binary component as compared to the
orbital angular momentum, such that the parameter $\xi =\chi _{1}\left(
L_{N}/S_{1}\right) \leq 0.1$. This latter condition selects the mass ratios $%
1:30$ and below. Similar considerations lead to the conclusion, that the
second spin can be neglected. Therefore, from all available angular momenta
of the system, the larger spin dominates. Hence we call the corresponding
waveforms spin-dominated waveforms (SDW).

The SDW were given as a double expansion in the PN parameter $%
\varepsilon $ and the smallness parameter $\xi $. The angle $\beta _{1}$
between the dominant spin and total angular momentum is also small, of order 
$\xi $ (this is because the larger spin gives the main contribution to the
total angular momentum) The usual mass variables employed for gravitational
waveforms, the chirp mass $M_{chirp}=\left( m_{1}m_{2}\right)
^{3/5}m^{-1/5}=m\eta ^{3/5}$, and the mass ratio $\nu $ relate to the total
mass $m$ and $\xi $ as follows: 

\begin{eqnarray}
M_{chirp} &=&m\left[ \frac{\varepsilon ^{1/2}\xi }{\left( 1+\varepsilon
^{1/2}\xi \right) ^{2}}\right] ^{3/5}  \notag \\
&\approx &m\varepsilon ^{3/10}\xi ^{3/5}\left( 1-\frac{6}{5}\varepsilon
^{1/2}\xi \right) ~,
\end{eqnarray}%

\begin{equation*}
\nu =\varepsilon ^{1/2}\xi ~.
\end{equation*}

The limits of validity of these waveforms were studied, their typical
lengths in the instrument best sensitivity range were estimated for the
Advanced LIGO, the Einstein Telescope, the LAGRANGE and the LISA
gravitational wave detectors, respectively. While the SDW hold for any mass
ratio smaller than $1:30$, lower bounds for the mass ratios were derived
from the best sensitivity frequency range expected for these forthcoming
gravitational wave detectors. The mass ratio lower bounds are $1:140$ for
Advanced LIGO, $7\times 10^{-4}$ for the Einstein Telescope, $7\times
10^{-7} $ for the LAGRANGE and $7\times 10^{-9}$ for the LISA missions,
respectively.

The SDW amplitudes were presented to ($\varepsilon ^{3/2},\xi $) orders;
while the phase of the gravitational waves to ($\varepsilon ^{2},\xi $)
orders (omitting the highest order mixed terms). To this accuracy the
amplitude includes the (leading order) spin-orbit contributions. The phase
includes the (leading order) spin-orbit, the self-spin and the mass
quadrupole-mass monopole contributions. Note that expressing the mass
ratio in terms of $\xi $ shifts the usual $\varepsilon $ order, and this is
what happens with the spin-spin contributions, which would appear only at
higher orders. This is consistent with the smaller body's spin being
insignificant in the chosen parameter range in comparison with the larger
spin and orbital angular momentum.

A comparison showed that the derived SDW, when written with the same
level of detail as the waveforms given in Appendix A of Ref. \cite{ABFO},
are approximately 80\% shorter. This is due to the smaller number of
variables and most importantly, the second series expansion we employ. We
expect that the SDW will turn useful both in modeling gravitational waves
emitted by binary systems consisting of an astrophysical black hole with a
smaller black hole or a neutron star companion; and for supermassive black
hole encounters.

\section{Acknowledgements}

L\'{A}G was partially supported by COST Action MP0905 "Black Holes in a
Violent Universe." ZK was supported by OTKA\ Grant No. 100216. We acknowledge the support of the European Union / European Social Fund Grant T\'{A}MOP-4.2.2.C-11/1/KONV-2012-0010.

\appendix

\section{Explicit Spin-dominated waveforms\label{appA}}

We give below the various contributions to the waveform expressed in the
double expansion ($\varepsilon $, $\xi $), remembering that $\beta _{1}=%
\mathcal{O}\left( \xi \right) $. The ($\varepsilon ^{0}$, $\xi ^{0}\,$)
contributions are%
\begin{eqnarray}
4h_{+}^{0} &=&\textstyle\sum\limits_{+,-}\left[ \left( \sin ^{2}\theta
-2\right) \cos (2\phi _{n}\pm 2\psi )c_{1}^{\left( \pm 0\right) }\right.  
\notag \\
&&\left. -2\sin \kappa _{1}\sin 2\theta \sin (\phi _{n}\pm 2\psi )k^{\left(
\pm \right) }\right]   \notag \\
&&+6\sin ^{2}\kappa _{1}\sin ^{2}\theta \cos 2\psi ~, \\
2h_{\times }^{0} &=&\textstyle\sum\limits_{+,-}\left[ \cos \theta \sin
(2\phi _{n}\pm 2\psi )c_{1}^{\left( \pm 0\right) }\right.   \notag \\
&&\left. -2\sin \theta \sin \kappa _{1}\cos (\phi _{n}\pm 2\psi )k^{\left(
\pm \right) }\right] ~,
\end{eqnarray}%
while the ($\varepsilon ^{0}$, $\beta _{1}$) contributions read%
\begin{eqnarray}
2h_{+}^{0\beta SO} &=&\textstyle\sum\limits_{+,-}\left[ \sin 2\theta \sin
(\phi _{n}\pm 2\psi )c_{2}^{\left( \pm 0\right) }\right.   \notag \\
&&\left. +\sin \kappa _{1}\left( \sin ^{2}\theta -2\right) \cos (2\phi
_{n}\pm 2\psi )k^{\left( \pm \right) }\right]   \notag \\
&&-3\sin 2\kappa _{1}\sin ^{2}\theta \cos 2\psi ~, \\
h_{\times }^{0\beta SO} &=&\textstyle\sum\limits_{+,-}\left[ \cos \theta
\sin \kappa _{1}\sin (2\phi _{n}\pm 2\psi )k^{\left( \pm \right) }\right.  
\notag \\
&&\left. +\sin \theta \cos (\phi _{n}\pm 2\psi )c_{2}^{\left( \pm 0\right) }
\right] ~.
\end{eqnarray}%
\begin{table}[tbp]
\caption{The terms $a_{i}^{\left( \pm n\right) }$ in all PN orders,
\thinspace $n$ is the PN order, $i$ is the number of the constant in the
appropriate PN order.}
\begin{center}
\begin{tabular}{ccc}
\hline\hline
$n$ & $i$ & $a_{\left( i\right) }^{\left( \pm n\right) }$ \\ \hline
$0$ & $1$ & $\mp 2k^{\left( \pm \right) }$ \\ 
& $2$ & $\mp k^{\left( \pm \right) }$ \\ \hline
$0.5$ & $1$ & $4k^{\left( \pm \right) }\left( 6-\sin ^{2}\theta \right) $ \\ 
& $2$ & $4k^{\left( \pm \right) }$ \\ 
& $3$ & $\pm 2\left( 6-\sin ^{2}\theta \right) k^{\left( \pm \right) }$ \\ 
& $4$ & $12k^{\left( \pm \right) }$ \\ 
& $5$ & $\pm 2k^{\left( \pm \right) }(2\sin ^{2}\theta -3)$ \\ 
& $6$ & $-2k^{\left( \pm \right) }$ \\ 
& $7$ & $\mp 2c_{2}^{\left( \pm 0\right) }\left( 6-\sin ^{2}\theta \right) $
\\ 
& $8$ & $\pm 2k^{\left( \pm \right) }$ \\ 
& $9$ & $44-34\sin ^{2}\theta \pm 2\left( 5\sin ^{2}\theta -46\right) \cos
\kappa _{1}$ \\ 
& $10$ & $22 \pm 46 \cos \kappa_{1}$ \\ 
& $11$ & $-2k^{\left( \pm \right) }\left( 3-2\sin ^{2}\theta \right) $ \\ 
\hline
$1$ & $1$ & $\pm 8k^{\left( \pm \right) }$ \\ 
& $2$ & $6k^{\left( \pm \right) }\left( \sin ^{2}\theta +5\right) $ \\ 
& $3$ & $2k^{\left( \pm \right) }\left( 4-\sin ^{2}\theta \right) $ \\ 
& $4$ & $\pm 2k^{\left( \pm \right) }\left( 2\sin ^{4}\theta +11\sin
^{2}\theta -38\right) $ \\ 
& $5$ & $6k^{\left( \pm \right) }\left( 3\sin ^{2}\theta +5\right) $ \\ 
& $6$ & $\pm 2k^{\left( \pm \right) }\left( 4\sin ^{2}\theta +19\right) $ \\ 
& $7$ & $-2k^{\left( \pm \right) }\left( 3\sin ^{2}\theta -4\right) $ \\ 
& $8$ & $\pm 2k^{\left( \pm \right) }\left( 4-\sin ^{2}\theta \right) $ \\ 
& $9$ & $\pm 6k^{\left( \pm \right) }\left( 5+\sin ^{2}\theta \right) $ \\ 
& $10$ & $\mp 4k^{\left( \pm \right) }$ \\ 
& $11$ & $k^{\left( \pm \right) }\left( 22+29\sin ^{2}\theta -16\sin
^{4}\theta \right) $ \\ 
& $12$ & $2k^{\left( \pm \right) }$ \\ 
& $13$ & $\pm 6k^{\left( \pm \right) }\left( 3\sin ^{2}\theta +5\right) $ \\ 
& $14$ & $-k^{\left( \pm \right) }\left( 20\sin ^{2}\theta +11\right) $ \\ 
& $15$ & $\mp 2k^{\left( \pm \right) }\left( 3\sin ^{2}\theta -4\right) $ \\ 
\hline
$1.5$ & $1$ & $\pm 4k^{\left( \pm \right) }\left( \sin ^{2}\theta -6\right) $
\\ 
& $2$ & $\pm 4k^{\left( \pm \right) }\left( \sin ^{4}\theta +42\sin
^{2}\theta \!-\!166\right) $ \\ 
& $3$ & $16k^{\left( \pm \right) }$ \\ 
& $4$ & $8k^{\left( \pm \right) }\left( \sin ^{4}\theta +8\sin ^{2}\theta
-28\right) $ \\ 
& $5$ & $8k^{\left( \pm \right) }\left( -332+94\sin ^{2}\theta +\sin
^{4}\theta \right) $ \\ 
& $6$ & $\pm 8k^{\left( \pm \right) }\left( 38-42\sin ^{2}\theta -9\sin
^{4}\theta \right) $ \\ 
& $7$ & $-16k^{\left( \pm \right) }\left( 152-46\sin ^{2}\theta -9\sin
^{4}\theta \right) $ \\ 
& $8$ & $\pm 24k^{\left( \pm \right) }\left( 3\sin ^{2}\theta -10\right) $
\\ 
& $9$ & $-8k^{\left( \pm \right) }\left( 160-204\sin ^{2}\theta -63\sin
^{4}\theta \right) $ \\ 
& $10$ & $\pm 4k^{\left( \pm \right) }\left( 3-2\sin ^{2}\theta \right) $ \\ 
& $11$ & $-8k^{\left( \pm \right) }\left( 14+3\sin ^{2}\theta \right) $ \\ 
& $12$ & $-16k^{\left( \pm \right) }\left( 15\sin ^{2}\theta +76\right) $ \\ 
& $13$ & $-8k^{\left( \pm \right) }\left( 5\sin ^{2}\theta +166\right) $ \\ 
& $14$ & $-8k^{\left( \pm \right) }\left( 80+63\sin ^{2}\theta \right) $ \\ 
& $15$ & $\pm 4k^{\left( \pm \right) }\left( 166-125\sin ^{2}\theta -8\sin
^{4}\theta \right) $ \\ 
& $16$ & $\mp 8k^{\left( \pm \right) }\left( 38-61\sin ^{2}\theta -24\sin
^{4}\theta \right) $ \\ 
& $17$ & $\pm 8k^{\left( \pm \right) }\left( 5-4\sin ^{2}\theta \right) $ \\ 
\hline\hline
\end{tabular}%
\end{center}
\label{table01}
\end{table}
\begin{table}[tbp]
\caption{The terms $b_{\left( i\right) 0}^{\left( \pm n\right) }$ in all PN
orders, \thinspace $n$ is the PN order, $i$ is the number of the constant in
the appropriate PN order.}
\begin{center}
\begin{tabular}{ccc}
\hline\hline
$n$ & $i$ & $b_{\left( i\right) 0}^{\left( \pm n\right) }$ \\ \hline
$0$ & $1$ & $-1$ \\ 
& $2$ & $-2$ \\ \hline
$0.5$ & $1$ & $\pm \left( 46-5\sin ^{2}\theta \right) $ \\ 
& $2$ & $\pm 3$ \\ 
& $3$ & $2-3\sin ^{2}\theta $ \\ 
& $4$ & $\pm 23$ \\ 
& $5$ & $-\cos 2\theta $ \\ 
& $6$ & $\mp 4$ \\ 
& $7$ & $-$ \\ 
& $8$ & $+3$ \\ 
& $9$ & $-15\left( 2-3\sin ^{2}\theta \right) $ \\ 
& $10$ & $-15$ \\ 
& $11$ & $\mp 4\left( 3-2\sin ^{2}\theta \right) $ \\ \hline
$1$ & $1$ & $8$ \\ 
& $2$ & $\mp 18\pm 7\sin ^{2}\theta $ \\ 
& $3$ & $\mp 3\sin ^{2}\theta \pm 6$ \\ 
& $4$ & $-22-29\sin ^{2}\theta +16\sin ^{4}\theta $ \\ 
& $5$ & $\pm 26\sin ^{2}\theta \mp 18$ \\ 
& $6$ & $+11+20\sin ^{2}\theta $ \\ 
& $7$ & $\mp 6\sin ^{2}\theta \pm 6$ \\ 
& $8$ & $2\left( 11-5\sin ^{2}\theta \right) $ \\ 
& $9$ & $6\left( 7+9\sin ^{2}\theta \right) $ \\ 
& $10$ & $-11$ \\ 
& $11$ & $\mp 3\left( 8-20\sin ^{2}\theta +7\sin ^{4}\theta \right) $ \\ 
& $12$ & $\pm 3$ \\ 
& $13$ & $3\left( 19\sin ^{2}\theta +14\right) $ \\ 
& $14$ & $\pm 12\cos 2\theta $ \\ 
& $15$ & $22-21\sin ^{2}\theta $ \\ \hline
$1.5$ & $1$ & $5\sin ^{2}\theta -6$ \\ 
& $2$ & $18\sin ^{4}\theta +252\sin ^{2}\theta -188$ \\ 
& $3$ & $\pm 20$ \\ 
& $4$ & $\pm 9\sin ^{4}\theta \mp 90\sin ^{2}\theta \pm 56$ \\ 
& $5$ & $\mp 4\left( 1184-172\sin ^{2}\theta -7\sin ^{4}\theta \right) $ \\ 
& $6$ & $2\left( 46+48\sin ^{2}\theta -99\sin ^{4}\theta \right) $ \\ 
& $7$ & $\mp 12\left( 72+110\sin ^{2}\theta -63\sin ^{4}\theta \right) $ \\ 
& $8$ & $144\left( \sin ^{2}\theta -2\right) $ \\ 
& $9$ & $\mp 3\left( -204+406\sin ^{2}\theta -189\sin ^{4}\theta \right) $
\\ 
& $10$ & $3-4\sin ^{2}\theta $ \\ 
& $11$ & $\pm 28\mp 31\sin ^{2}\theta $ \\ 
& $12$ & $\mp \allowbreak 432\mp 876\sin ^{2}\theta $ \\ 
& $13$ & $\mp 4\left( 71\sin ^{2}\theta +592\right) $ \\ 
& $14$ & $\pm 306\mp 651\sin ^{2}\theta $ \\ 
& $15$ & $2\left( 94-173\sin ^{2}\theta -24\sin ^{4}\theta \right) $ \\ 
& $16$ & $-2\left( 46+25\sin ^{2}\theta -180\sin ^{4}\theta \right) $ \\ 
& $17$ & $48\cos ^{2}\theta $ \\ \hline\hline
\end{tabular}%
\end{center}
\label{table02}
\end{table}
\begin{table}[tbp]
\caption{The terms $d_{\left( i\right) 0}^{\left( \pm n\right) }$ in 0.5, 1
and 1.5 PN orders, \thinspace $n$ is the PN order, $i$ is the number of the
constant in the appropriate PN order.}
\begin{center}
\begin{tabular}{ccc}
\hline\hline
$n$ & $i$ & $d_{\left( i\right) 0}^{\left( \pm n\right) }$ \\ \hline
$0.5$ & $1$ & $5\left( 3\sin ^{2}\theta -2\right) $ \\ 
& $2$ & $-1$ \\ 
& $3$ & $-$ \\ 
& $4$ & $-5$ \\ 
& $5$ & $-$ \\ 
& $6$ & $3$ \\ 
& $7$ & $-3\left( 2-3\sin ^{2}\theta \right) $ \\ 
& $8$ & $-$ \\ 
& $9$ & $-$ \\ 
& $10$ & $-$ \\ 
& $11$ & $3\cos 2\theta $ \\ \hline
$1$ & $1$ & $\mp 4$ \\ 
& $2$ & $6-14\sin ^{2}\theta $ \\ 
& $3$ & $2\left( \sin ^{2}\theta -1\right) $ \\ 
& $4$ & $\mp 2\left( 8-20\sin ^{2}\theta +7\sin ^{4}\theta \right) $ \\ 
& $5$ & $6-7\sin ^{2}\theta $ \\ 
& $6$ & $\mp 16\sin ^{2}\theta \pm 8$ \\ 
& $7$ & $3\sin ^{2}\theta -2$ \\ 
& $8$ & $\pm 9\left( \sin ^{2}\theta -2\right) $ \\ 
& $9$ & $\pm 3\left( 18-7\sin ^{2}\theta \right) $ \\ 
& $10$ & $\pm 9$ \\ 
& $11$ & $4\left( 2-8\sin ^{2}\theta +7\sin ^{4}\theta \right) $ \\ 
& $12$ & $-2$ \\ 
& $13$ & $\pm 6\left( 9-13\sin ^{2}\theta \right) $ \\ 
& $14$ & $2\left( 7\sin ^{2}\theta -2\right) $ \\ 
& $15$ & $\mp 18\cos ^{2}\theta $ \\ \hline
$1.5$ & $1$ & $-$ \\ 
& $2$ & $\mp 6\sin ^{4}\theta \pm 72\sin ^{2}\theta \mp 20$ \\ 
& $3$ & $-12$ \\ 
& $4$ & $-15\sin ^{4}\theta +22\sin ^{2}\theta -8$ \\ 
& $5$ & $1920-2832\sin ^{2}\theta -84\sin ^{4}\theta $ \\ 
& $6$ & $\pm 6\left( 10-44\sin ^{2}\theta +27\sin ^{4}\theta \right) $ \\ 
& $7$ & $-4\left( 88-422\sin ^{2}\theta +171\sin ^{4}\theta \right) $ \\ 
& $8$ & $\pm 12\left( 14-9\sin ^{2}\theta \right) $ \\ 
& $9$ & $-9\left( 28-126\sin ^{2}\theta +105\sin ^{4}\theta \right) $ \\ 
& $10$ & $-$ \\ 
& $11$ & $9\sin ^{2}\theta -4$ \\ 
& $12$ & $-176+756\sin ^{2}\theta $ \\ 
& $13$ & $12\left( 7\sin ^{2}\theta +80\right) $ \\ 
& $14$ & $-126+\allowbreak 189\sin ^{2}\theta $ \\ 
& $15$ & $\pm 2\left( 10-41\sin ^{2}\theta +36\sin ^{4}\theta \right) $ \\ 
& $16$ & $\mp 6\left( 10-49\sin ^{2}\theta +44\sin ^{4}\theta \right) $ \\ 
& $17$ & $\mp 4\left( 7-8\sin ^{2}\theta \right) $ \\ \hline\hline
\end{tabular}%
\end{center}
\label{table03}
\end{table}
\begin{table}[tbp]
\caption{The terms $b_{\left( i\right) 1}^{\left( \pm n\right) }$ in 1 and
1.5 PN orders, \thinspace $n$ is the PN order, $i$ is the number of the
constant in the appropriate PN order.}
\begin{center}
\begin{tabular}{ccc}
\hline\hline
$n$ & $i$ & $b_{\left( i\right) 1}^{\left( \pm n\right) }$ \\ \hline
$1$ & $1$ & $-1$ \\ 
& $2$ & $-$ \\ 
& $3$ & $-$ \\ 
& $4$ & $-2\left( 2-8\sin ^{2}\theta +7\sin ^{4}\theta \right) $ \\ 
& $5$ & $-$ \\ 
& $6$ & $2-7\sin ^{2}\theta $ \\ 
& $7$ & $-$ \\ 
& $8$ & $-8\cos ^{2}\theta $ \\ 
& $9$ & $8\left( 3-7\sin ^{2}\theta \right) $ \\ 
& $10$ & $4$ \\ 
& $11$ & $-$ \\ 
& $12$ & $-$ \\ 
& $13$ & $-4\left( 7\sin ^{2}\theta -6\right) $ \\ 
& $14$ & $-$ \\ 
& $15$ & $-4\left( 2-3\sin ^{2}\theta \right) $ \\ \hline
$1.5$ & $1$ & $-$ \\ 
& $2$ & $-15\sin ^{4}\theta +12\sin ^{2}\theta -2$ \\ 
& $3$ & $\mp 5$ \\ 
& $4$ & $-$ \\ 
& $5$ & $\mp \left( 236-294\sin ^{2}\theta +21\sin ^{4}\theta \right) $ \\ 
& $6$ & $3\left( 6-36\sin ^{2}\theta +45\sin ^{4}\theta \right) $ \\ 
& $7$ & $\mp 3\left( 232-510\sin ^{2}\theta +243\sin ^{4}\theta \right) $ \\ 
& $8$ & $9\left( 6-5\sin ^{2}\theta \right) $ \\ 
& $9$ & $-$ \\ 
& $10$ & $-$ \\ 
& $11$ & $-$ \\ 
& $12$ & $\mp \allowbreak 348\pm 591\sin ^{2}\theta $ \\ 
& $13$ & $\pm \left( 273\sin ^{2}\theta -118\right) $ \\ 
& $14$ & $-$ \\ 
& $15$ & $\left( 2-13\sin ^{2}\theta +12\sin ^{4}\theta \right) $ \\ 
& $16$ & $-9\left( 2-13\sin ^{2}\theta +12\sin ^{4}\theta \right) $ \\ 
& $17$ & $-3\left( 3-4\sin ^{2}\theta \right) $ \\ \hline\hline
\end{tabular}%
\end{center}
\label{table04}
\end{table}
\begin{table}[tbp]
\caption{The terms $d_{\left( i\right) 1}^{\left( \pm n\right) }$ in 1.5 PN
order, \thinspace $n$ is the PN order, $i$ is the number of the constant in
the appropriate PN order.}
\begin{center}
\begin{tabular}{ccc}
\hline\hline
$n$ & $i$ & $d_{\left( i\right) 1}^{\left( \pm n\right) }$ \\ 
$1.5$ & $1$ & $-$ \\ 
& $2$ & $-$ \\ 
& $3$ & $1$ \\ 
& $4$ & $-$ \\ 
& $5$ & $28-126\sin ^{2}\theta +105\sin ^{4}\theta $ \\ 
& $6$ & $-$ \\ 
& $7$ & $27\left( 8-22\sin ^{2}\theta +15\sin ^{4}\theta \right) $ \\ 
& $8$ & $-$ \\ 
& $9$ & $-$ \\ 
& $10$ & $-$ \\ 
& $11$ & $-$ \\ 
& $12$ & $-243\sin ^{2}\theta +108$ \\ 
& $13$ & $7\left( 2-3\sin ^{2}\theta \right) $ \\ 
& $14$ & $-$ \\ 
& $15$ & $-$ \\ 
& $16$ & $-$ \\ 
& $17$ & $-$ \\ \hline\hline
\end{tabular}%
\end{center}
\label{table05}
\end{table}
The ($\varepsilon ^{0.5}$, $\xi ^{0}\,$) corrections to the gravitational
waveforms $h_{_{\times }^{+}}$ are

\begin{eqnarray}
64h_{+}^{0.5} &=&4\cos \theta \sin \kappa _{1}\sin ^{2}\theta \left[ -45\sin
^{2}\kappa _{1}\cos 3\psi \right.  \notag \\
&&\left. +\cos \psi \left( 5\sin ^{2}\kappa _{1}-4\right) \right]  \notag \\
&&+\textstyle\sum\limits_{+,-}\left\{ -\sin ^{2}\kappa _{1}\sin \theta \left[
\left( \sin ^{2}\theta \right. \right. \right.  \notag \\
&&\left. -2\right) \left( \sin (3\phi _{n}\pm \psi )k^{\left( \pm \right)
}\right)  \notag \\
&&\left. -45\left( 2-3\sin ^{2}\theta \right) \left( \sin (\phi _{n}\pm
3\psi )k^{\left( \pm \right) }\right) \right]  \notag \\
&&+\sin \theta \left[ \left( \sin (\phi _{n}\pm \psi )c_{1}^{\left( \pm
0.5\right) }\right) \right.  \notag \\
&&\left. -9\left( \sin ^{2}\theta -2\right) \left( \sin (3\phi _{n}\pm 3\psi
)c_{2}^{\left( \pm 0.5\right) }\right) \right]  \notag \\
&&+2\cos \theta \sin \kappa _{1}\left[ \left( \cos (2\phi _{n}\pm \psi
)c_{3}^{\left( \pm 0.5\right) }\right) \right. +9\left( 2\right.  \notag \\
&&\left. \left. \left. -3\sin ^{2}\theta \right) \left( \cos (2\phi _{n}\pm
3\psi )c_{1}^{\left( \pm 0\right) }\right) \right] ~\right\} ~,
\end{eqnarray}%

\begin{eqnarray}
32h_{\times }^{0.5} &=&-16\sin 2\kappa _{1}\sin \psi \sin ^{2}\theta ~ 
\notag \\
&&+\textstyle\sum\limits_{+,-}\left\{ \cos \theta \sin ^{2}\kappa _{1}\sin
\theta \left[ \cos (3\phi _{n}\pm \psi )k^{\left( \pm \right) }\right.
\right.  \notag \\
&&\left. +45\left( \cos (\phi _{n}\pm 3\psi )k^{\left( \pm \right) }\right) 
\right]  \notag \\
&&+\frac{1}{2}\sin 2\theta \left[ \left( \cos (\phi _{n}\pm \psi
)c_{4}^{\left( \pm 0.5\right) }\right) \right.  \notag \\
&&\left. +9\left( \cos (3\phi _{n}\pm 3\psi )c_{2}^{\left( \pm 0.5\right)
}\right) \right]  \notag \\
&&+2\sin \kappa _{1}\left[ \left( \sin (2\phi _{n}\pm \psi )c_{5}^{\left(
\pm 0.5\right) }\right) \right.  \notag \\
&&\left. \left. -9\cos 2\theta \left( \sin (2\phi _{n}\pm 3\psi
)c_{1}^{\left( \pm 0\right) }\right) \right] \right\} ~,
\end{eqnarray}%
while the ($\varepsilon ^{0.5}$, $\beta _{1}$) contributions read 

\begin{eqnarray}
64h_{+}^{0.5\beta } &=&4\cos \kappa _{1}\cos \theta \sin ^{2}\theta \left[
135\cos \left( 3\psi \right) \sin ^{2}\kappa _{1}\right.  \notag \\
&&\left. +\cos \psi \left( 4-15\sin ^{2}\kappa _{1}\right) \right] +%
\textstyle\sum\limits_{+,-}\left\{ 2\cos \theta \left[ 9\left( 2\right.
\right. \right.  \notag \\
&&\left. -3\sin ^{2}\theta \right) \left( \cos (2\phi _{n}\pm 3\psi
)c_{6}^{\left( \pm 0.5\right) }\right)  \notag \\
&&\left. +\cos (2\phi _{n}\pm \psi )c_{7}^{\left( \pm 0.5\right) }\right]
-\sin \kappa _{1}\sin \theta \left( \sin ^{2}\theta \right.  \notag \\
&&\left. -2\right) \left[ 27\left( \sin (3\phi _{n}\pm 3\psi )c_{1}^{\left(
\pm 0\right) }\right) \right.  \notag \\
&&\left. +\sin (3\phi _{n}\pm \psi )c_{8}^{\left( \pm 0.5\right) }\right]
+\sin \kappa _{1}\sin \theta \left[ 45\left( 2\right. \right.  \notag \\
&&\left. -3\sin ^{2}\theta \right) \left( \sin (\phi _{n}\pm 3\psi
)c_{8}^{\left( \pm 0.5\right) }\right)  \notag \\
&&\left. \left. +\sin (\phi _{n}\pm \psi )c_{9}^{\left( \pm 0.5\right) } 
\right] \right\} ~,
\end{eqnarray}%
\begin{eqnarray}
32h_{\times }^{0.5\beta } &=&32\sin \psi \sin ^{2}\theta \cos 2\kappa _{1} 
\notag \\
&&+\textstyle\sum\limits_{+,-}\left\{ \cos \theta \sin \kappa _{1}\sin
\theta \left[ 27\cos (3\phi _{n}\pm 3\psi )c_{1}^{\left( \pm 0\right)
}\right. \right.  \notag \\
&&+\left[ \cos (3\phi _{n}\pm \psi )+45\cos (\phi _{n}\pm 3\psi )\right]
c_{8}^{\left( \pm 0.5\right) }  \notag \\
&&\left. +c_{10}^{\left( \pm 0.5\right) }\cos (\phi _{n}\pm \psi )\right] 
\notag \\
&&-\left[ 18\cos 2\theta \left( \sin (2\phi _{n}\pm 3\psi )c_{6}^{\pm
}\right) \right.  \notag \\
&&\left. \left. -2\sin (2\phi _{n}\pm \psi )c_{11}^{\left( \pm 0.5\right) } 
\right] ~\right\} ~.
\end{eqnarray}%
The ($\varepsilon ^{1}$, $\xi ^{0}\,$) corrections to the gravitational
waveforms $h_{_{\times }^{+}}$ are

\begin{eqnarray}
48h_{+}^{1} &=&2\sin ^{2}\theta \sin ^{2}\kappa _{1}\left[ \sin ^{2}\kappa
_{1}5\left( 7\sin ^{2}\theta \right. \right.  \notag \\
&&\left. -6\right) \left( \cos 2\psi -4\cos 4\psi \right)  \notag \\
&&\left. -2\left( 15\sin ^{2}\theta +51\right) \cos 2\psi \right]  \notag \\
&&+\textstyle\sum\limits_{+,-}\left\{ 16\sin ^{3}\kappa _{1}\sin 2\theta
\left( 7\sin ^{2}\theta \right. \right.  \notag \\
&&\left. -3\right) \sin (\phi _{n}\pm 4\psi )k^{\left( \pm \right) }-\left(
\sin ^{2}\theta \right.  \notag \\
&&\left. -2\right) \sin ^{2}\kappa _{1}\sin ^{2}\theta \cos (4\phi _{n}\pm
2\psi )c_{1}^{\left( \pm 0\right) }  \notag \\
&&+4\left( \sin ^{2}\theta -2\right) \sin ^{2}\theta \cos (4\phi _{n}\pm
4\psi )c_{1}^{\left( \pm 1\right) }  \notag \\
&&+2\sin \kappa _{1}\sin 2\theta \left[ \sin (\phi _{n}\pm 2\psi
)c_{2}^{\left( \pm 1\right) }\right.  \notag \\
&&+\sin (3\phi _{n}\pm 2\psi )c_{3}^{\left( \pm 1\right) }  \notag \\
&&\left. -8\cos ^{2}\theta \sin (3\phi _{n}\pm 4\psi )c_{2}^{\left( \pm
0.5\right) }\right]  \notag \\
&&+2\cos (2\phi _{n}\pm 2\psi )c_{4}^{\left( \pm 1\right) }  \notag \\
&&-16\sin ^{2}\kappa _{1}\left[ 7\sin ^{4}\theta -2\left( 4\sin ^{2}\theta
\right. \right.  \notag \\
&&\left. \left. \left. -1\right) \right] \cos (2\phi _{n}\pm 4\psi
)c_{1}^{\left( \pm 0\right) }\right\} ~,
\end{eqnarray}%

\begin{eqnarray}
24h_{\times }^{1} &=&60\cos \kappa _{1}\cos \theta \sin ^{2}\kappa _{1}\sin
2\psi \sin ^{2}\theta  \notag \\
&&+\textstyle\sum\limits_{+,-}\left[ 8\sin ^{3}\kappa _{1}\sin \theta \left(
7\sin ^{2}\theta \right. \right.  \notag \\
&&\left. -6\right) \cos (\phi _{n}\pm 4\psi )k^{\left( \pm \right) }  \notag
\\
&&-\cos \theta \sin ^{2}\kappa _{1}\sin ^{2}\theta \sin (4\phi _{n}\pm 2\psi
)c_{1}^{\left( \pm 0\right) }  \notag \\
&&+2\sin \kappa _{1}\sin \theta \cos (\phi _{n}\pm 2\psi )c_{5}^{\left( \pm
1\right) }  \notag \\
&&+4\cos \theta \sin ^{2}\theta \sin (4\phi _{n}\pm 4\psi )c_{1}^{\left( \pm
1\right) }  \notag \\
&&+2\cos \theta \sin (2\phi _{n}\pm 2\psi )c_{6}^{\left( \pm 1\right) } 
\notag \\
&&-8\cos \theta \sin ^{2}\kappa _{1}\left( 7\sin ^{2}\theta -2\right) \sin
(2\phi _{n}\pm 4\psi )c_{1}^{\left( \pm 0\right) }  \notag \\
&&-8\left( 2-3\sin ^{2}\theta \right) \sin \kappa _{1}\sin \theta \cos
(3\phi _{n}\pm 4\psi )c_{2}^{\left( \pm 0.5\right) }  \notag \\
&&\left. +2\sin \kappa _{1}\sin \theta \cos (3\phi _{n}\pm 2\psi
)c_{7}^{\left( \pm 1\right) }\right] ~,
\end{eqnarray}%
while the ($\varepsilon ^{1}$, $\beta _{1}$) contributions read 

\begin{eqnarray}
24h_{+}^{1\beta } &=&\cos \kappa _{1}\sin \kappa _{1}\sin ^{2}\theta \left[
4\left( 15\sin ^{2}\theta +51\right) \cos 2\psi \right.  \notag \\
&&\left. +20\sin ^{2}\kappa _{1}\left( 7\sin ^{2}\theta -6\right) \left(
4\cos 4\psi -\cos 2\psi \right) \right]  \notag \\
&&+\textstyle\sum\limits_{+,-}\left\{ \sin 2\theta \sin (3\phi _{n}\pm 2\psi
)c_{8}^{\left( \pm 1\right) }\right.  \notag \\
&&+\sin 2\theta \sin (\phi _{n}\pm 2\psi )c_{9}^{\left( \pm 1\right) } 
\notag \\
&&+8\cos ^{2}\theta \sin 2\theta \sin (3\phi _{n}\pm 4\psi )c_{10}^{\left(
\pm 1\right) }  \notag \\
&&+2\sin \kappa _{1}\cos (2\phi _{n}\pm 2\psi )c_{11}^{\left( \pm 1\right) }
\notag \\
&&+8\sin ^{2}\kappa _{1}\sin 2\theta \left( 7\sin ^{2}\theta \right.  \notag
\\
&&\left. -3\right) \sin (\phi _{n}\pm 4\psi )\left[ \pm 3k^{\left( \pm
\right) }+4\sin ^{2}\kappa _{1}\right]  \notag \\
&&+16\sin \kappa _{1}\left( 2-8\sin ^{2}\theta \right.  \notag \\
&&\left. +7\sin ^{4}\theta \right) \cos (2\phi _{n}\pm 4\psi )c_{12}^{\left(
\pm 1\right) }  \notag \\
&&+\sin \kappa _{1}\sin ^{2}\theta \left( \sin ^{2}\theta -2\right) \left[
\cos (4\phi _{n}\pm 2\psi )c_{12}^{\left( \pm 1\right) }\right.  \notag \\
&&\left. \left. -8\cos (4\phi _{n}\pm 4\psi )c_{2}^{\left( \pm 0.5\right) } 
\right] \right\} ~,
\end{eqnarray}%
\begin{eqnarray}
12h_{\times }^{1\beta } &=&30\cos \theta \sin \kappa _{1}\sin 2\psi \sin
^{2}\theta \left( 3\sin ^{2}\kappa _{1}\right.  \notag \\
&&\left. -2\right) +\textstyle\sum\limits_{+,-}\left\{ \sin \theta \cos
(\phi _{n}\pm 2\psi )c_{13}^{\left( \pm 1\right) }\right.  \notag \\
&&+2\cos \theta \sin \kappa _{1}\left[ \sin (2\phi _{n}\pm 2\psi
)c_{14}^{\left( \pm 1\right) }\right.  \notag \\
&&\left. +4\left( 7\sin ^{2}\theta -2\right) \sin (2\phi _{n}\pm 4\psi
)c_{12}^{\left( \pm 1\right) }\right]  \notag \\
&&+\cos \theta \sin \kappa _{1}\sin ^{2}\theta \left[ \sin (4\phi _{n}\pm
2\psi )c_{12}^{\left( \pm 1\right) }\right.  \notag \\
&&\left. -8\sin (4\phi _{n}\pm 4\psi )c_{2}^{\left( \pm 0.5\right) }\right] 
\notag \\
&&+\sin \theta \left[ \cos (3\phi _{n}\pm 2\psi )c_{15}^{\left( \pm 1\right)
}\right.  \notag \\
&&\left. +4\left( 2-3\sin ^{2}\theta \right) \cos (3\phi _{n}\pm 4\psi
)c_{10}^{\left( \pm 1\right) }\right]  \notag \\
&&+4\sin ^{2}\kappa _{1}\sin \theta \left( 6-7\sin ^{2}\theta \right) \left(
\mp 3k^{\left( \pm \right) }\right.  \notag \\
&&\left. \left. -4\sin ^{2}\kappa _{1}\right) \cos (\phi _{n}\pm 4\psi
)\right\} ~.
\end{eqnarray}%
The ($\varepsilon ^{1}$, $\xi ^{0}\,$) and ($\varepsilon ^{1}$, $\xi \,$) SO
corrections to the gravitational waveforms $h_{_{\times }^{+}}$ are%
\begin{equation}
h_{+}^{1SO}=\frac{\chi _{1}}{2}\sin \theta \textstyle\sum\limits_{+,-}\left[
\pm k^{\left( \pm \right) }\sin (\phi _{n}\pm \psi )\right] ~,
\end{equation}%

\begin{eqnarray}
h_{\times }^{1SO} &=&\frac{\chi _{1}}{2}\sin \theta \left\{ 2\sin \kappa
_{1}\sin \,\psi \sin \theta \right.  \notag \\
&&\left. +\cos \theta \textstyle\sum\limits_{+,-}\left[ \pm k^{\left( \pm
\right) }\cos (\phi _{n}\pm \psi )\right] \right\} ~,
\end{eqnarray}%

\begin{eqnarray}
\frac{2}{\chi _{1}}h_{+}^{1\beta SO} &=&\textstyle\sum\limits_{+,-}\left\{
\mp k^{\pm }\cos \theta \sin (2\phi _{n}\pm \psi -\phi _{1})\right.  \notag
\\
&&+\sin \kappa _{1}\sin \theta \left[ \pm \sin (\phi _{n}\pm \psi )\right. 
\notag \\
&&\left. \left. \pm \cos (\phi _{n}\pm \psi -\phi _{1})\right] \right\} ~.
\end{eqnarray}%

\begin{eqnarray}
\frac{4}{\chi _{1}}h_{\times }^{1\beta SO} &=&-\left[ 2\left( \sin \phi
_{1}+2\right) \cos \kappa _{1}\sin \psi \right.  \notag \\
&&\left. +2\cos \phi _{1}\cos \psi \right] \sin ^{2}\theta  \notag \\
&&+\textstyle\sum\limits_{+,-}\left\{ 2\cos \theta \sin \kappa _{1}\sin
\theta \left[ \pm \cos (\phi _{n}\pm \psi )\right. \right.  \notag \\
&&\left. \mp \sin (\phi _{n}\pm \psi -\phi _{1})\right] \pm \left( \sin
^{2}\theta \right.  \notag \\
&&\left. \left. -2\right) k^{\pm }\cos (2\phi _{n}\pm \psi -\phi
_{1})\right\} ~,
\end{eqnarray}%
The ($\varepsilon ^{1.5}$, $\xi ^{0}\,$) corrections to the gravitational
waveforms $h_{_{\times }^{+}}$ are 
\begin{eqnarray}
12288h_{+}^{1.5} &=&12\cos \theta \sin \kappa _{1}\sin ^{2}\theta \left\{
\cos 3\psi \left[ 1701\left( 2\right. \right. \right.  \notag \\
&&\left. -3\sin ^{2}\theta \right) \sin ^{4}\kappa _{1}+72\sin ^{2}\kappa
_{1}\left( 63\sin ^{2}\theta \right.  \notag \\
&&\left. \left. +178\right) \right] +\cos \psi \left[ -14\left( 2-3\sin
^{2}\theta \right) \sin ^{4}\kappa _{1}\right.  \notag \\
&&\left. -8\sin ^{2}\kappa _{1}\left( 7\sin ^{2}\theta +162\right) +16\left(
\sin ^{2}\theta +66\right) \right]  \notag \\
&&\left. -4375\left( 2-3\sin ^{2}\theta \right) \sin ^{4}\kappa _{1}\cos
5\psi \right\} +\textstyle\sum\limits_{+,-}\left[ 2\left( \sin ^{2}\theta
\right. \right.  \notag \\
&&\left. -2\right) \sin ^{4}\kappa _{1}\sin ^{3}\theta k^{\left( \pm \right)
}\sin (5\phi _{n}\pm \psi )  \notag \\
&&+4\cos \theta \sin ^{3}\kappa _{1}\sin ^{2}\theta \cos (4\phi _{n}\pm \psi
)c_{1}^{\left( \pm 1.5\right) }  \notag \\
&&+16\cos \theta \sin \kappa _{1}\cos (2\phi _{n}\pm \psi )c_{2}^{\left( \pm
1.5\right) }  \notag \\
&&+1250\sin ^{4}\kappa _{1}\sin \theta \left( 105\sin ^{4}\theta \right. 
\notag \\
&&\left. -126\sin ^{2}\theta +28\right) k^{\left( \pm \right) }\sin (\phi
_{n}\pm 5\psi )  \notag \\
&&+625\left( \sin ^{2}\theta -2\right) \sin ^{3}\theta \sin (5\phi _{n}\pm
5\psi )c_{3}^{\left( \pm 1.5\right) }  \notag \\
&&+6\sin ^{2}\kappa _{1}\sin \theta \sin (3\phi _{n}\pm \psi )c_{4}^{\left(
\pm 1.5\right) }  \notag \\
&&+243\left( \sin ^{2}\theta -2\right) \sin ^{2}\kappa _{1}\sin ^{3}\theta
\sin (5\phi _{n}\pm 3\psi )c_{2}^{\left( \pm 0.5\right) }  \notag \\
&&+4\sin \theta \sin (\phi _{n}\pm \psi )c_{5}^{\left( \pm 1.5\right) } 
\notag \\
&&+5000\cos \theta \sin ^{3}\kappa _{1}\left( 15\sin ^{4}\theta -12\sin
^{2}\theta \right.  \notag \\
&&\left. +2\right) \cos (2\phi _{n}\pm 5\psi )c_{1}^{\left( \pm 0\right) } 
\notag \\
&&-1250\cos \theta \sin \kappa _{1}\sin ^{2}\theta \left( 5\sin ^{2}\theta
\right.  \notag \\
&&\left. -6\right) \cos (4\phi _{n}\pm 5\psi )c_{1}^{\left( \pm 1\right) } 
\notag \\
&&+1875\sin ^{2}\kappa _{1}\sin \theta \left( 8-22\sin ^{2}\theta \right. 
\notag \\
&&\left. +15\sin ^{4}\theta \right) \sin (3\phi _{n}\pm 5\psi )c_{2}^{\left(
\pm 0.5\right) }  \notag \\
&&+216\cos \theta \sin \kappa _{1}\cos (2\phi _{n}\pm 3\psi )c_{6}^{\left(
\pm 1.5\right) }  \notag \\
&&+27\sin \theta \sin (3\phi _{n}\pm 3\psi )c_{7}^{\left( \pm 1.5\right) } 
\notag \\
&&+54\cos \theta \sin \kappa _{1}\sin ^{2}\theta \cos (4\phi _{n}\pm 3\psi
)c_{8}^{\left( \pm 1.5\right) }  \notag \\
&&\left. +54\sin ^{2}\kappa _{1}\sin \theta \sin (\phi _{n}\pm 3\psi
)c_{9}^{\left( \pm 1.5\right) }\right] ~,
\end{eqnarray}%
\begin{eqnarray}
6144h_{\times }^{1.5} &=&192\cos \kappa _{1}\sin \kappa _{1}\sin ^{2}\theta 
\left[ \sin \psi \left( 64\right. \right.  \notag \\
&&\left. -\sin ^{2}\kappa _{1}\left( 7\sin ^{2}\theta -6\right) +4\sin
^{2}\theta \right)  \notag \\
&&\left. +27\sin 3\psi \sin ^{2}\kappa _{1}\left( 7\sin ^{2}\theta -6\right) 
\right]  \notag \\
&&+\textstyle\sum\limits_{+,-}\left\{ +4\sin ^{3}\kappa _{1}\sin ^{2}\theta
\sin (4\phi _{n}\pm \psi )c_{10}^{\left( \pm 1.5\right) }\right.  \notag \\
&&-2\cos \theta \sin ^{4}\kappa _{1}\sin ^{3}\theta \cos (5\phi _{n}\pm \psi
)k^{\left( \pm \right) }  \notag \\
&&-243\cos \theta \sin ^{2}\kappa _{1}\sin ^{3}\theta \cos (5\phi _{n}\pm
3\psi )c_{2}^{\left( \pm 0.5\right) }  \notag \\
&&-625\cos \theta \sin ^{3}\theta \cos (5\phi _{n}\pm 5\psi )c_{3}^{\left(
\pm 1.5\right) }  \notag \\
&&+3\sin ^{2}\kappa _{1}\sin 2\theta \cos (3\phi _{n}\pm \psi
)c_{11}^{\left( \pm 1.5\right) }  \notag \\
&&+27\cos \theta \sin \theta \cos (3\phi _{n}\pm 3\psi )c_{12}^{\left( \pm
1.5\right) }  \notag \\
&&+1875\cos \theta \sin ^{2}\kappa _{1}\sin \theta \left( 4\right.  \notag \\
&&\left. -9\sin ^{2}\theta \right) \cos (3\phi _{n}\pm 5\psi )c_{2}^{\left(
\pm 0.5\right) }  \notag \\
&&+2\sin 2\theta \cos (\phi _{n}\pm \psi )c_{13}^{\left( \pm 1.5\right) } 
\notag \\
&&+27\sin ^{2}\kappa _{1}\sin 2\theta \cos (\phi _{n}\pm 3\psi
)c_{14}^{\left( \pm 1.5\right) }  \notag \\
&&+8\sin \kappa _{1}\sin (2\phi _{n}\pm \psi )c_{15}^{\left( \pm 1.5\right)
}+4375\left( 2\right.  \notag \\
&&\left. -3\sin ^{2}\theta \right) \sin ^{4}\kappa _{1}\sin 2\theta
k^{\left( \pm \right) }\cos (\phi _{n}\pm 5\psi )  \notag \\
&&+108\sin \kappa _{1}\sin (2\phi _{n}\pm 3\psi )c_{16}^{\left( \pm
1.5\right) }  \notag \\
&&+162\sin \kappa _{1}\sin ^{2}\theta \sin (4\phi _{n}\pm 3\psi
)c_{17}^{\left( \pm 1.5\right) }  \notag \\
&&-2500\sin ^{3}\kappa _{1}\left( 2-13\sin ^{2}\theta \right.  \notag \\
&&\left. +12\sin ^{4}\theta \right) \sin (2\phi _{n}\pm 5\psi )c_{1}^{\left(
\pm 0\right) }  \notag \\
&&-1250\sin \kappa _{1}\sin ^{2}\theta \left( 3\right.  \notag \\
&&\left. \left. -4\sin ^{2}\theta \right) \sin (4\phi _{n}\pm 5\psi
)c_{1}^{\left( \pm 1\right) }\right\} ~,
\end{eqnarray}%
The ($\varepsilon ^{1.5}$, $\xi ^{0}\,$) SO corrections to the gravitational
waveforms $h_{_{\times }^{+}}$ are%
\begin{eqnarray}
\frac{2}{\chi _{1}}h_{+}^{1.5SO} &=&4\sin \kappa _{1}\left[ \cos \kappa
_{1}\sin \kappa _{1}\cos 2\phi _{n}\right.  \notag \\
&&-\cos 2\kappa _{1}\cos \theta \sin \phi _{n}\sin \theta  \notag \\
&&+\cos \kappa _{1}\sin \kappa _{1}\sin ^{2}\theta \left( 6\sin ^{2}\psi
\right.  \notag \\
&&\left. \left. -2+\sin ^{2}\phi _{n}\right) \right]  \notag \\
&&+\textstyle\sum\limits_{+,-}\left\{ 2\cos \theta \sin \kappa _{1}\sin
\theta \left[ \left( \mp 3k^{\left( \pm \right) }\right. \right. \right. 
\notag \\
&&\left. \left. -4\sin ^{2}\kappa _{1}\right) \sin (\phi _{n}\pm 2\psi ) 
\right]  \notag \\
&&+\cos (2\phi _{n}\pm 2\psi )\left[ -2k^{\left( \pm \right) }+\left( 2\cos
\kappa _{1}\right. \right.  \notag \\
&&\left. \left. \left. \mp 3\right) \sin ^{2}\kappa _{1}\right] \left( \sin
^{2}\theta -2\right) \right\} ~,
\end{eqnarray}%
\begin{eqnarray}
\frac{1}{\chi _{1}}h_{\times }^{1.5SO} &=&-2\cos \phi _{n}\sin \kappa
_{1}\left( \sin \theta \cos 2\kappa _{1}\right.  \notag \\
&&\left. +\cos \theta \sin 2\kappa _{1}\sin \phi _{n}\right)  \notag \\
&&+\textstyle\sum\limits_{+,-}\left\{ \cos \theta \sin (2\phi _{n}\pm 2\psi )%
\left[ -2k^{\left( \pm \right) }\right. \right.  \notag \\
&&\left. +\left( 2\cos \kappa _{1}\mp 3\right) \sin ^{2}\kappa _{1}\right] 
\notag \\
&&+\sin \kappa _{1}\sin \theta \left[ \mp 3k^{\left( \pm \right) }\right. 
\notag \\
&&\left. \left. -4\sin ^{2}\kappa _{1}\right] \cos (\phi _{n}\pm 2\psi
)\right\} ~,
\end{eqnarray}%
The ($\varepsilon ^{1.5}$, $\xi ^{0}\,$) tail corrections to the
gravitational waveforms $h_{_{\times }^{+}}$ are%
\begin{eqnarray}
\frac{2}{\pi }h_{+}^{1.5tail}\! &=&\!6\sin ^{2}\kappa _{1}\sin ^{2}\theta
\cos 2\psi  \notag \\
&&+\textstyle\sum\limits_{+,-}\left[ \cos (2\phi _{n}\pm 2\psi
)c_{1}^{\left( \pm 0\right) }\left( \sin ^{2}\theta -2\right) \right.  \notag
\\
&&\left. -2k^{\left( \pm \right) }\sin \kappa _{1}\sin 2\theta \sin (\phi
_{n}\pm 2\psi )\right] ~,
\end{eqnarray}%
\begin{eqnarray}
\frac{1}{\pi }h_{\times }^{1.5tail}\!\! &=&\!\!\textstyle\sum\limits_{+,-}%
\left[ \cos \theta \sin (2\phi _{n}\pm 2\psi )c_{1}^{\left( \pm 0\right)
}\right.  \notag \\
&&\left. -2\sin \kappa _{1}\sin \theta k^{\left( \pm \right) }\cos (\phi
_{n}\pm 2\psi )\right] ~,
\end{eqnarray}

In the above expressions we have introduced the notations $k^{\left( \pm
\right) }=\cos \kappa _{1}\mp 1$. The constant coefficients\ $c_{i}^{\left(
\pm n\right) }$~, also appearing in the various PN orders of $h_{_{\times
}^{+}}$ are structured as 
\begin{eqnarray}
c_{\left( i\right) }^{\left( \pm n\right) } &=&a_{\left( i\right) }^{\left(
\pm n\right) }+\sin ^{2}\kappa _{1}\sum_{j=0}^{1}\left( b_{\left( i\right)
j}^{\left( \pm n\right) }\right.   \notag \\
&&\left. +d_{\left( i\right) j}^{\left( \pm n\right) }\cos \kappa
_{i}\right) \sin ^{2j}\kappa _{1}~.  \label{coeffstruct}
\end{eqnarray}%
Here \thinspace $n$ denotes the PN order of the coefficients and $i$, $j$
are serial numbers. The contributions $a_{\left( i\right) }^{\left( \pm
n\right) }$, $b_{\left( i\right) 0}^{\left( \pm n\right) }$, $d_{\left(
i\right) 0}^{\left( \pm n\right) }$,$~b_{\left( i\right) 1}^{\left( \pm
n\right) }$ and $d_{\left( i\right) 1}^{\left( \pm n\right) }$ are given in
the Tables \ref{table01}, \ref{table02}, \ref{table03}, \ref{table04} and %
\ref{table05}. Note that in the nonprecessing case ($\kappa _{1}=0,\pi $)
the contributions $a_{\left( i\right) }^{\left( \pm n\right) }$ (with $%
k^{+}=0$, $k^{-}=2$) fully define the coefficients $c_{i}^{\left( \pm
n\right) }$.


\begin{thebibliography}{99}
\bibitem{plunge} R. Sturani, S. Fischetti, L. Cadonati, G. M. Guidi, J.
Healy, D. Shoemaker, A. Vicer\'{e}, \textit{J.Phys.Conf.Ser.} \textbf{243}
012007, (2010) E-print: arXiv:1005.0551; F. Pretorius, \textit{Class. Quant.
Grav.} \textbf{22} 425 (2005); M. Campanelli, C. O. Lousto, Y. Zlochower, 
\textit{Phys. Rev.} D \textbf{74}, 041501 (2006); M. Campanelli, C. O.
Lousto, Y. Zlochower, \textit{Phys. Rev.} D \textbf{74}, 084023 (2006); J.
G. Baker, J. Centrella, D. Choi, M. Koppitz, J. van Meter, \textit{Phys. Rev.%
} D \textbf{73 }104002, (2006).

\bibitem{ringdown} E. Berti, V. Cardoso, A. O. Starinets,\textit{\ Class.
Quantum Grav.} \textbf{26}, 163001 (2009), E-print: arXiv:0905.2975.

\bibitem{BOC} B. M. Barker, R. F. O'Connell, \textit{Phys. Rev.} D \textbf{12%
}, 329 (1975).

\bibitem{BOC2} B. M. Barker. R. F. O'Connell, \textit{Gen. Relativ. Gravit.} 
\textbf{2}, 1428 (1979).

\bibitem{KWW} L. E. Kidder, C. M. Will, A. G. Wiseman, \textit{Phys. Rev.} D 
\textbf{47}, R4183 (1993).

\bibitem{KIDDER} L. E. Kidder, \textit{Phys. Rev.} D \textbf{52}, 821 (1995).

\bibitem{ACST} T. A. Apostolatos, C. Cutler, G. J. Sussman, K. S. Thorne, 
\textit{Phys.Rev.} D \textbf{49}, 6274 (1994).

\bibitem{SO} F. D. Ryan, \textit{Phys. Rev.} D \textbf{53}, 3064 (1996); R.
Rieth, G. Sch\"{a}fer, \textit{Class. Quantum Grav.} \textbf{14}, 2357
(1997); L. \'{A}. Gergely, Z. Perj\'{e}s, M. Vas\'{u}th,\textit{\ Phys. Rev.}
D \textbf{57}, 876 (1998); R. F. O'Connell, \textit{Phys. Rev. Letters} 
\textbf{93}, 081103 (2004); C. M. Will, \textit{Phys. Rev.} D \textbf{71},
084027 (2005); J. Zeng, C. M. Will, \textit{Gen. Rel. Grav.} \textbf{39}
1661 (2007); L. \'{A}. Gergely, Z. I. Perj\'{e}s, M. Vas\'{u}th, \textit{%
Phys. Rev.} D \textbf{58}, 124001(1998).

\bibitem{SpinFlip1} L. \'{A}. Gergely, P. L. Biermann, \textit{Astrophys. J.}
\textbf{697}, 1621 (2009).

\bibitem{SpinFlip2} L. \'{A}. Gergely, P. L. Biermann, L. I. Caramete, 
\textit{Class. Quantum Grav.} \textbf{27}, 194009 (2010).

\bibitem{SS} T. A. Apostolatos,\textit{\ Phys. Rev.} D \textbf{52}, 605
(1995); T. A. Apostolatos, \textit{Phys. Rev.} D \textbf{54}, 2438 (1996);
H. Wang, C. M. Will, \textit{Phys. Rev.} D \textbf{75}, 064017 (2007); J. Maj%
\'{a}r, \textit{Phys. Rev.} D \textbf{80}, 104028 (2009); A. Klein, Ph.
Jetzer, \textit{Phys. Rev.} D 81 124001 (2010), E-print: arXiv:1005.2046.

\bibitem{spinspin1} L. \'{A}. Gergely, \textit{Phys. Rev.} D \textbf{61},
024035 (1999).

\bibitem{spinspin2} L. \'{A}. Gergely, \textit{Phys. Rev.} D \textbf{62},
024007 (2000).

\bibitem{ABFO} K. G. Arun, A. Buonanno, G. Faye, E. Ochsner, \textit{Phys.
Rev.} D \textbf{79}, 104023 (2009).

\bibitem{Poisson} E. Poisson, \textit{Phys. Rev.} D \textbf{57}, 5287 (1998).

\bibitem{quadrupol} L. \'{A}. Gergely, Z. Keresztes, \textit{Phys. Rev.} D 
\textbf{67}, 024020 (2003).

\bibitem{QM} E. E. Flanagan, T. Hinderer, \textit{Phys. Rev.} D \textbf{75},
124007 (2007), E-print: arXiv:0704.0389; \'{E}. Racine, \textit{Phys. Rev.}
D \textbf{78}, 044021 (2008).

\bibitem{BFB} G. Faye, L. Blanchet, A. Buonanno, \textit{Phys. Rev.} D 
\textbf{74,} 104033 (2006); L. Blanchet, A. Buonanno, G. Faye, \textit{Phys.
Rev.} D \textbf{74}, 104034 (2006); Erratum-ibid. D \textbf{75} 049903,
(2007); Erratum-ibid. D \textbf{81} 089901, (2010).

\bibitem{Renorm} L. \'{A}. Gergely, B. Mik\'{o}czi, \textit{Phys. Rev.} D 
\textbf{79}, 064023 (2009); L. \'{A}. Gergely, P. L. Biermann, B. Mik\'{o}%
czi, Z. Keresztes, \textit{Class. Quantum Grav.} \textbf{26}, 204006 (2009).

\bibitem{Inspiral1} L. \'{A}. Gergely, \textit{Phys. Rev.} D \textbf{81},
084025 (2010).

\bibitem{Inspiral2} L. \'{A}. Gergely, \textit{Phys. Rev. }D \textbf{82,}
104031 (2010), E-print: arXiv:1005.5330.

\bibitem{MajarVasuth1} J. Maj\'{a}r, M. Vas\'{u}th, \textit{Phys. Rev.} D 
\textbf{77} 104005 (2008).

\bibitem{CornishKey} N. J. Cornish, J. S. Key, \textit{Phys. Rev.} D \textbf{%
82} 044028 (2010).

\bibitem{multipoles} R. A. Porto, A. Ross, I. Z. Rothstein E-print:
arXiv:1203.2962 (2012).

\bibitem{OV} J. R. Oppenheimer, G. M. Volkoff, \textit{Phys. Rev.} \textbf{55%
}, 374 (1939).

\bibitem{Bombacci} I. Bombaci, \textit{Astron. Astrophys.} \textbf{305}, 871
(1996).

\bibitem{MassSpin} L. \'{A}. Gergely, P. L. Biermann, E-print:
arXiv:1208.5251 qr-qc (2012).

\bibitem{LIGO1} the LIGO Scientific Collaboration, the Virgo Collaboration, 
\textit{Phys. Rev.} D \textbf{85}, 082002 (2012).

\bibitem{LIGO2} the LIGO Scientific Collaboration, E-print: arXiv:1201.5999.

\bibitem{LIGO3} B. Abadie et al. (LIGO Scientific Collaboration \& Virgo) 
\textit{Phys. Rev.} D \textbf{83} 122005 \ arXiv:1102.3781 [gr-qc].\ (2011).

\bibitem{aLIGO} G. M. Harry (for the LIGO Scientific Collaboration) \textit{%
Class. Quantum Grav.} \textbf{27} 084006 (2010).

\bibitem{LISA} K. G. Arun et al., \textit{Class. Quantum Grav.} \textbf{26} 094027 (2009).

\bibitem{eLISA} P. Amaro-Seoane et al. arXiv:1201.3621v1.

\bibitem{LAGRANGE} J. W. Conklin, et. al., arXiv:1111.5264 (2011).

\bibitem{ET} J. R. Gair, I. Mandel, M. C. Miller, M. Volonteri,\ \textit{%
Gen. Relativ. Gravit.} \textbf{43} 485 (2011).

\bibitem{BCV2} A. Buonanno, Y. Chen, M. Vallisneri, \textit{Phys. Rev.} D 
\textbf{67} 104025 (2003); Erratum-ibid. D \textbf{74} 029904 (2006).

\bibitem{PBCV} Y. Pan, A. Buonanno, Y. Chen nad M. Vallisneri, \textit{Phys.
Rev.} D \textbf{69} 104017 (2004); Erratum-ibid. D \textbf{74} 029905 (2006).

\bibitem{omegadot} B. Mik\'{o}czi, M. Vas\'{u}th, L. \'{A}. Gergely, \textit{%
Phys. Rev}. D \textbf{71}, 124043 (2005).

\bibitem{LEVIN} J. Levin, S. T. McWilliams, H. Contreras, \textit{Class.
Quant. Grav.} \textbf{28} 175001 (2011).

\bibitem{BIWW} L. Blanchet, B. R. Iyer, C. M. Will, A. G. Wiseman \textit{%
Class.Quant.Grav.} \textbf{13} 575 (1996).
\end{thebibliography}
\end{document}